\documentclass[12pt]{article}

\usepackage{amsmath}
\usepackage{graphicx}
\usepackage{bm}
\usepackage{amssymb}
\usepackage{amsfonts}
\usepackage{comment}
\usepackage{mathtools}
\usepackage[numbers, sort&compress]{natbib}
\usepackage{xcolor}
\usepackage{url}
\usepackage{tcolorbox}
\usepackage[colorlinks,linktocpage,linkcolor=cyan,citecolor=cyan]{hyperref}
\usepackage[top=2.5cm, bottom=2.5cm, left=2.5cm, right=2.5cm, headheight=14pt, headsep=0.8cm, footskip=1.0cm]{geometry}
\usepackage{newtxtext, newtxmath}
\usepackage{microtype}

\def\l{\left}
\def\r{\right}
\def\ddd{\mathrm{d}}
\def\nn{\nonumber}
\def\bcc#1\ecc{\begin{tcolorbox}[center, colback = gray!10!white, colframe=white, width=17cm] #1 \end{tcolorbox}\noindent}

\title{Evolution of Linear Perturbations under Time-Dependent Hubble Friction I: SR-USR-SR Inflation}

\author{
	Wen Li, \textsuperscript{1} \quad
	Chao Chen \textsuperscript{1,}\thanks{Corresponding author: cchao012@just.edu.cn;}
	\\
	\small \textsuperscript{1} School of Science, Jiangsu University of Science and Technology,	Zhenjiang, 212100, China
	}

\date{}

\begin{document}
	
\maketitle

\begin{abstract}
	The origin of the finite dip in the curvature power spectrum of instantaneous Slow-Roll (SR)-Ulta-Slow-Roll (USR)-SR inflation remains controversial at linear order, and its full spectral features still lack a complete asymptotic analytical description. We revisit linear perturbation dynamics in this framework. Using the junction method and asymptotic expansions of Hankel functions, we for the first time derive accurate and simple asymptotic expressions for mode evolution and the resulting power spectrum, based on three systematic rules for dominant-term identification across transitions. We find the finite dip arises from cancellation between two growing modes within linear perturbation theory, rather than between constant and growing terms as previously suggested. We also provide analytical descriptions of the amplitude enhancement and the two oscillatory patterns observed in the spectrum. All asypmtotic analytical results are validated by numerical calculations with the reconstructed SR-USR-SR inflationary potential.
\end{abstract}

\tableofcontents

\section{Introduction}

The slow-roll (SR) inflation, the simplest and most popular inflationary scenario, naturally explains the nearly scale-invariant power spectrum of primordial curvature perturbations observed on large scales in the cosmic microwave background radiation (CMB)~\cite{Planck:2018jri}. 
However, the statistics of primordial curvature perturbations on small scales remain poorly constrained (c.f., Fig.~14 in Ref.~\cite{Escriva:2022duf} or Fig.~1 in Ref.~\cite{Green:2020jor}), primarily due to the challenges of highly nonlinear evolutions and the limited sensitivity of current experiments. 
This leaves significant scope for theoretical models beyond the vanilla slow-roll scenario, particularly those that predict a strong enhancement of small-scale curvature perturbations. In this context, the ultra-slow-roll (USR) inflation stands as the most prominent scenario, enabling such an enhancement that leads to rich phenomena, including primordial black hole formation and scalar-induced gravitational waves (e.g., see reviews~\cite{Carr:2016drx, Yuan:2019fwv, Carr:2020xqk, Chang:2020tji, Yuan:2021qgz, Domenech:2021ztg, Escriva:2022duf, Ozsoy:2023ryl, Pi:2024lsu} for more details).
The USR inflation offers not only rich phenomenology but also compelling theoretical motivations: it serves as a simple model for analyzing non-perturbative effects on primordial curvature perturbation statistics~\cite{Biagetti:2018pjj,Pattison:2018bct,Cruces:2018cvq,Firouzjahi:2018vet,Atal:2019cdz,Ezquiaga:2019ftu,Figueroa:2020jkf,Figueroa:2021zah, Pattison:2021oen,Biagetti:2021eep,Davies:2021loj,Cai:2021zsp,Cai:2022erk,Pi:2022ysn,Hooshangi:2022lao,Ahmadi:2022lsm,Hooshangi:2023kss,Raatikainen:2023bzk,Tomberg:2023kli,Inui:2024sce,Inui:2024fgk,Wang:2024nmd,Escriva:2025ftp,Caravano:2025diq,Nassiri-Rad:2025dsa,Firouzjahi:2025ipu}, while the quantum one-loop corrections induced by enhanced small-scale perturbations during this phase remain an open and intriguing topic~\cite{Kristiano:2022maq,  Choudhury:2023vuj,Kristiano:2023scm,Riotto:2023gpm,Choudhury:2023rks,Firouzjahi:2023aum,Firouzjahi:2023ahg,Firouzjahi:2023btw,Franciolini:2023agm,Cheng:2023ikq,Fumagalli:2023zzl,Fumagalli:2023loc,Maity:2023qzw,Tada:2023rgp,Firouzjahi:2023bkt,Davies:2023hhn,Iacconi:2023ggt,Firouzjahi:2024psd,Inomata:2024lud,Ballesteros:2024zdp,Fumagalli:2024jzz,Caravano:2024moy,Sheikhahmadi:2024peu,Firouzjahi:2024sce,Inomata:2025bqw,Fang:2025vhi,Inomata:2025pqa,Braglia:2025qrb,Kristiano:2025ajj,Ema:2025ftj,Fang:2025kgf,Iacconi:2026uzo}.

\begin{figure}[h]
	\centering
	\includegraphics[width=0.38\textwidth]{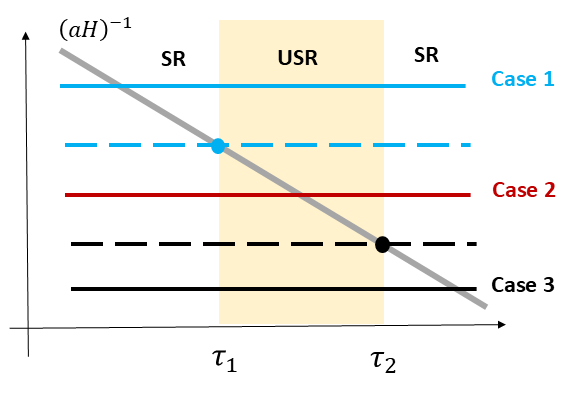}
	\raisebox{0.5\height}{\includegraphics[width=0.38\textwidth]{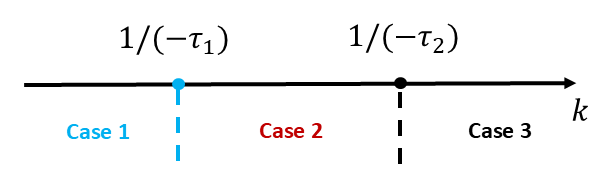}}
	\caption{
		A sketch illustrating three representative $k$ regimes of the mode function $\chi_{k}(\tau)$, in relation to SR-USR-SR instantaneous transitions denoted by $\tau_{1}$ and $\tau_{2}$.
	}
	\label{fig:k_regimes}
\end{figure}

The simplest USR model exhibits a sandwich structure: where the USR phase serves as an intermediate stage between two SR phases, as illustrated in the left panel of Fig.~\ref{fig:k_regimes}. Throughout this paper, the USR phase is defined by the second slow-roll parameter, $\eta \equiv \dot{\epsilon}/ (\epsilon H) = -6$,
\footnote{
	In literature, the definition of the USR phase varies. For instance, Refs.~\cite{Tsamis:2003px,Kinney:2005vj,Namjoo:2012aa} characterize it by a flat potential, i.e., $V_{,\phi} \equiv \ddd V/\ddd\phi = 0$, while Refs.~\cite{Martin:2012pe, Motohashi:2014ppa,Pattison:2018bct,Motohashi:2019rhu,Motohashi:2025qgd} define USR as a regime where $\beta \equiv \ddot{\phi}/(H \dot{\phi}) = -3$.
	In this work, we adopt the definition used in Refs.~\cite{Dimopoulos:2017ged,Germani:2017bcs}, which aligns with other conventions through the relation $\eta = -6 - 2 {V_{,\phi} \over H \dot{\phi}} + 2\epsilon = 2 \beta + 2 \epsilon$.
}
where $\epsilon \equiv - \dot{H}/H^2$ is the slow-roll parameter and the dot refers to the time derivative.
The USR inflation can be classified within a broader concept of constant-roll (CR) inflation, where $\eta$ is a constant~\cite{Motohashi:2025qgd}.
During the USR phase, the inflationary potential becomes nearly flat, and the inflaton’s motion is dominated by Hubble friction rather than the potential gradient. This makes the USR phase a non-attractor regime, where perturbations exhibit distinct behaviors compared with that of SR phase.
In particular, the linear comoving curvature perturbation $\mathcal{R}$ develops a growing mode on superhorizon scales during USR~\cite{Dimopoulos:2017ged,Byrnes:2018txb, Carrilho:2019oqg,Ozsoy:2019lyy,Tasinato:2020vdk,Cheng:2021lif,Ng:2021hll,Cole:2022xqc,Karam:2022nym,Gu:2023mmd,Zhao:2024yzg}, which ultimately enhances the curvature power spectrum $\mathcal{P}_{\mathcal{R}}(k) \equiv {k^3 \over 2\pi^2} |\mathcal{R}_{k}(\tau_{\rm end})|^2$ at the end of inflation $\tau_{\rm end}\rightarrow 0^{-}$.
\footnote{
	Here, the superscript ``-'' of $0$ means that $\tau_{\rm end}$ approaches $0$ from the left.
}
This enhancement can be understood intuitively from the linear equation of motion (EoM) for $\mathcal{R}_k$,
\begin{equation} \label{eq:EoM_Rk}
	\ddot{\mathcal{R}}_k + (3 + \eta) H \dot{\mathcal{R}}_k + {k^2 \over a^2} \mathcal{R}_k = 0 ~,
\end{equation}
where a growing mode emerges in the solution, once the effective Hubble friction coefficient $(3 + \eta) H$ becomes negative (i.e., when $\eta < -3$)~\cite{Byrnes:2018txb,Dimopoulos:2017ged}. 
Utilizing the junction method, Ref.~\cite{Byrnes:2018txb} first identified the steepest possible $k^4$ growth in the linear power spectrum $\mathcal{P}_{\mathcal{R}}(k)$ in USR inflation. 
Later work~\cite{Carrilho:2019oqg} showed that an even steeper scaling, $k^5 (\ln k)^2$, can arise if a CR phase with $\eta = -1$ is incorporated.
This work also employed the perturbative method to analyze the dip structure observed in numerical linear power spectra~\cite{Garcia-Bellido:2017mdw, Byrnes:2018txb, Carrilho:2019oqg}. 
In addition, a steeper $k^6$ growth is achievable in models featuring a varying sound speed~\cite{Zhai:2023azx} and for $\alpha$-vacuum initial state~\cite{Cielo:2024poz}. Furthermore, it has been demonstrated that scenarios involving multiple transitions during inflation can lead to significantly stronger enhancements~\cite{Tasinato:2020vdk}, a $k^8$ maximum slope can be realized in the presence of two non-attractor phases.

Additionally, a characteristic imprint of the SR-USR transition is the dip structure in the linear power spectrum. This non-trivial feature emerges on the large scales, making it a prominent and observationally accessible signature for testing the USR phase or other non-attractor dynamics~\cite{Cielo:2024poz}.
At leading order, Ref.~\cite{Carrilho:2019oqg} argued that the dip of $\mathcal{P}_{\mathcal{R}}(k)$ is zero, arising from the exact cancellation between the constant mode and a negative growing mode of $\mathcal{R}_{k}$. This explanation, however, contradicts numerical results that show a non-zero dip. 
Recently, Ref.~\cite{Fujita:2025imc} employed the $\delta N$ formalism to analyze the dip structure and the non-linearity in presence of non-attractor phase(s), demonstrating that the conjugate field momentum is crucial for the dip structure. Reference~\cite{Briaud:2025hra} developed a transfer-matrix formalism to study the linear curvature power spectrum in the presence of non-attractor phase(s), showing that the amplitude at the dip always scales as the inverse square root of the peak amplitude.

In this paper, we revisit linear curvature power spectrum in the conventional SR-USR-SR inflationary framework with instantaneous transitions. Our analysis is confined strictly to linear tree-level perturbation theory. Within this well-defined theoretical boundary, we aim to resolve long-standing ambiguities about the origin and full spectral structure of the curvature power spectrum in this minimal scenario.
Adopting the junction method and asymptotic expansions of Hankel functions as in Refs~\cite{Byrnes:2018txb, Carrilho:2019oqg}, we show that by consistently retaining the dominant contributions, the finite dip in $\mathcal{P}_{\mathcal{R}}(k)$ can be naturally explained within linear perturbation theory without requiring multiple or smooth transitions. Moreover, we demonstrate that this dip arises from the cancellation between two growing terms in the power spectrum. 
Additionally, for the first time, we derive appropriate asymptotic expressions that capture the time evolution of power spectra across SR–USR–SR transitions over various $k$ regimes, as illustrated in Fig.~\ref{fig:k_regimes}. These simple asymptotic analytical results are straightforward to compute theoretically and provide robust experimental tests, helping impose more reliable constraints on USR inflation~\cite{Byrnes:2018txb, Fujita:2025imc}.

The key to deriving accurate asymptotic expressions for the linear curvature power spectrum lies in systematically identifying the dominant contributions.
To do so, we propose three simple systematic rules to guide the construction of asymptotic expansions across transitions:
\begin{tcolorbox}[center, colback= gray!10!white, colframe= black, width= 12cm]
\begin{itemize}
	\item Rule 1: Identify the dominant terms at each transition;
	\item Rule 2: Identify the dominant terms at later times;
	\item Rule 3: Rule 1 must be applied prior to Rule 2.
\end{itemize}
\end{tcolorbox}
\noindent
Notice that, failure to follow Rule 3 may result in the incorrect neglect of dominant terms near transitions, leading to an inaccurate description of the mode function's time evolution.
As an example, Eq.~\eqref{case1:rule1} will demonstrate that in Case 1 of Fig.~\ref{fig:k_regimes}, a higher-order term in the constant mode can dominate over growing modes at the first SR–USR transition, ultimately producing a finite dip in the subsequent evolution, a subtlety that has largely been overlooked in previous studies.

We begin with a general scenario described by the following EoM for the mode function $\chi_k$,
\begin{equation} \label{eq:eom}
	\ddot{\chi}_k + 3 \tilde{H}(t) \dot{\chi}_k + {k^2 \over a^2} \chi_k = 0 ~,
\end{equation}
where the Hubble friction coefficient $\tilde{H}(t)$ has definite asymptotic behaviors: $\tilde{H}(t) \rightarrow \text{positive constants}$ in both the early-time limit $t \rightarrow 0$ and the late-time limit $t \rightarrow \infty$. 
The mode $\chi_k$ may represent the linear comoving curvature perturbation (c.f., Eq.~\eqref{eq:EoM_Rk}) or the spectator field perturbation during inflation~\cite{Pi:2021dft,Meng:2022low, Chen:2023lou}. 
The key challenge is to provide a clear analytical understanding of the resulting power spectrum $\mathcal{P}_{\chi}(\tau_{\rm end},k) = {k^3 \over 2\pi^2} |\chi_{k}(\tau_{\rm end})|^2$ at the end of inflation. Addressing this challenge hinges on obtaining accurate approximations for the full time evolution of the mode function $\chi_{k}(\tau)$.

This paper is organized as follows. In Sec.~\ref{sec:analytical}, we review the Hankel function solutions to the EoM~\eqref{eq:eom} under the quasi-de Sitter approximation. We then introduce the junction method for treating instantaneous transitions and examine the superhorizon and subhorizon asymptotics of Hankel functions, which enable a simple yet precise approximation for the power spectrum $\mathcal{P}_{\chi}(\tau,k)$. Subsequently, Sec.~\ref{sec:asymptotic} provides a detailed analysis of the time evolution of $\mathcal{P}_{\chi}(\tau,k)$ for the three representative cases of $k$ regimes illustrated in Fig.~\ref{fig:k_regimes}. Based on the results, we derive analytical expressions for the final power spectrum at the end of inflation. These expressions allow us to explain the characteristic features observed in the spectrum, including the dip, $k^4$-growth and two oscillation patterns. In Appendix~\ref{app:check}, we perform a consistent verification of the asymptotic expansions for the power spectra during both the SR and USR phases, utilizing the explicit analytical form of the Hankel functions. In Appendix~\ref{app:potential}, we adopt the Hamilton-Jacobi formalism to reconstruct the instantaneous SR-USR-SR inflationary potential (c.f., Eq.~\eqref{eq:final_potential}), which is then used to numerically compute the power spectrum and verify the validity of our analytical results.

\section{Analytic Solutions from the Junction Method}\label{sec:analytical}

\subsection{General solutions}

Applying the quasi-de Sitter approximation $aH = - {1\over\tau (1 - \epsilon)}$ (i.e., $\epsilon$ a constant) to Eq.~\eqref{eq:eom}, we obtain 
\begin{equation} \label{eq:chi_eom}
	\chi_{k}'' + { 1 - 2d \over \tau} \chi_{k}' + k^2 \chi_{k} = 0 ~,
\end{equation}
where the prime refers to the derivatives with respect to the conformal time $\ddd\tau = \ddd t/a(t)$.
Its solution for a constant $d \equiv {3 h - \epsilon \over 2 (1 - \epsilon)}$, with $h \equiv \tilde{H}/H$, is given by
\begin{equation} \label{eq:chi_sol_AB}
	\chi_{k}(\tau) = A (-\tau)^{d} H_{d}^{(1)}(- k \tau) + B (-\tau)^{d} H_{d}^{(2)}(- k \tau) ~,
\end{equation}
where the two unknown coefficients $A$ and $B$ can be determined by the initial conditions of $\chi_{k}(\tau)$, or the junction method as we will discuss later.

The main task of this paper is to analyze the asymptotic behaviors of the general solution~\eqref{eq:chi_sol_AB} across SR-USR-SR transitions, as illustrated in the left panel of Fig.~\ref{fig:k_regimes}.
The thick gray backslash denotes the comoving Hubble horizon during inflation, while the solid horizontal lines in blue, red and black represent three distinct $k$ regimes of the mode function $\chi_{k}(\tau)$, corresponding to the classification in the right panel of Fig.~\ref{fig:k_regimes}. 
The blue dashed line can be interpreted as the boundary of Case 1, corresponding to the $k$ mode that exits the horizon at the SR-USR transition $\tau_{1}$. Similarly, the black dashed line is regarded as the limit of Case 3, such that the $k$ mode exits the horizon at the USR-SR transition $\tau_{2}$.

\subsection{Junction method: SR-USR-SR}

One can approximate the non-trivial evolution of the inflationary background as a series of instantaneous transitions with different values of $d$, and then match perturbations using smooth junction conditions. This approach, known as the junction method, has been widely applied to the approximation of linear perturbations, in particular the enhancement of linear perturbations~\cite{Byrnes:2018txb,Carrilho:2019oqg,Tasinato:2020vdk,Fu:2022ssq,Pi:2022zxs,Wang:2024vfv,Wang:2024wxq,Fujita:2025imc}.

First, we consider a general transition between two phases (also see appendixes in Refs.~\cite{Byrnes:2018txb, Fu:2022ssq}), i.e., $d_1 \rightarrow d_2$, which have the following two solutions according to Eq.~\eqref{eq:chi_sol_AB},
\begin{subequations} \label{eq:sol1}
	\begin{align}
		\chi_{1, k}(\tau) &= A_1 (-\tau)^{d_1} H_{d_1}^{(1)}(- k \tau) + B_1 (-\tau)^{d_1} H_{d_1}^{(2)}(- k \tau) ~,
		\\ \label{eq:sol2}
		\chi_{2, k}(\tau) &= A_2 (-\tau)^{d_2} H_{d_2}^{(1)}(- k \tau) + B_2 (-\tau)^{d_2} H_{d_2}^{(2)}(- k \tau) ~,
	\end{align}
\end{subequations}
respectively. Using the junction conditions~\cite{Deruelle:1995kd,Byrnes:2018txb,Carrilho:2019oqg}, 
\begin{equation} \label{eq:junction_condition}
	\chi_{1, k}(\tau_1) = \chi_{2, k}(\tau_1) ~,
	\quad
	\chi_{1, k}'(\tau_1) = \chi_{2, k}'(\tau_1) ~,
\end{equation}
with $\tau_1$ the transition moment, we can solve the coefficients $\{ A_2, B_2\}$ in terms of $\{ A_1, B_1 \}$ as
\begin{subequations} \label{eq:coeff_A2B2}
		\begin{align}
			A_2 &=
			\frac{1}{4} i \pi A_1 k (-\tau_1)^{d_1-d_2+1} \l[ H_{d_1}^{(1)}(-k \tau_1) H_{d_2-1}^{(2)}(-k \tau_1)-H_{d_1-1}^{(1)}(-k \tau_1) H_{d_2}^{(2)}(-k\tau_1) \r]
			\nn\\&\quad
			+ \frac{1}{4} i \pi  B_1 k (-\tau_1)^{d_1-d_2+1} \l[ H_{d_1}^{(2)}(-k\tau_1) H_{d_2-1}^{(2)}(-k \tau_1)-H_{d_1-1}^{(2)}(-k \tau_1) H_{d_2}^{(2)}(-k \tau_1) \r] ~,
			\\
			B_2 
			&= \frac{1}{4} i \pi  A_1 k (-\tau_1)^{d_1-d_2 +1} \l[ H_{d_1-1}^{(1)}(-k \tau_1) H_{d_2}^{(1)}(-k \tau_1) - H_{d_1}^{(1)}(-k\tau_1) H_{d_2-1}^{(1)}(-k \tau_1) \r]
			\nn\\&\quad
			+ \frac{1}{4} i \pi B_1 k (-\tau_1)^{d_1-d_2+1} \l[ H_{d_1-1}^{(2)}(-k\tau_1) H_{d_2}^{(1)}(-k \tau_1)-H_{d_1}^{(2)}(-k \tau_1) H_{d_2-1}^{(1)}(-k\tau_1) \r] ~.
		\end{align}
\end{subequations}
It is clearly that $A_2, B_2$ are functions of $k$. One can iteratively apply the same process and get the solutions for each successive phases, maintaining the same structure of expressions in Eq.~\eqref{eq:coeff_A2B2}.

In this paper, we assume the inflation to start from a SR phase and Bunch-Davies initial vacuum state. Applying the de Sitter approximation $\epsilon =0$, we derive the SR solution based on the general solution~\eqref{eq:chi_sol_AB} as $\chi_{k}^{\rm SR1}(\tau) = { H \over \sqrt{2 k}} e^{- i k\tau}  \l( {i \over k} - \tau \r)$, i.e., $h=1$, $A = -H \sqrt{\pi}/2$ and $B = 0$. Its power spectrum is thus given as
\begin{equation} \label{eq:sr1_full}
	\mathcal{P}_{\chi}^{\rm SR1}(\tau, k) 
	= {k^3 \over2\pi^2} |\chi_{k}^{\rm SR1}(\tau)|^2 
	= \mathcal{P}_{\chi,{\rm CMB}} \l[ 1 + (-k\tau)^2 \r] ~,
\end{equation}
where
\begin{equation}
	\mathcal{P}_{\chi,{\rm CMB}} \equiv {H^2 \over 4 \pi^2} ~.
\end{equation}
Inside the horizon ($- k\tau \gg 1$), the second term dominates, we obtain
\begin{equation} \label{eq:sr1_sub}
	\mathcal{P}_{\chi}^{\rm SR1}(\tau, k) \simeq \mathcal{P}_{\chi,{\rm CMB}} (-k\tau)^2 ~.
\end{equation}
At some fixed time, e.g., the end of inflation $\tau_{\rm end}$, $\mathcal{P}_{\chi}^{\rm SR1}(\tau_{\rm end}, k) \propto k^2$ for the subhorizon modes.
On superhorizon scales ($- k\tau \ll 1$), we obtain
\begin{equation} \label{eq:sr1_super}
	\mathcal{P}_{\chi}^{\rm SR1}(\tau, k) \simeq \mathcal{P}_{\chi,{\rm CMB}} ~,
\end{equation}
which is the famous constant mode of superhorizon linear perturbations.

Based on solutions in Eq.~\eqref{eq:coeff_A2B2}, we consider the transition from SR to USR, such that $d_1=3/2$, $d_2 = -3/2$. Also, it is sufficient to take the de Sitter limit $\epsilon = 0$ to the SR solution for our purpose. We obtain
\begin{subequations} \label{eq:A2B2}
	\begin{align}
		A_2 &= -\frac{3 \sqrt{\pi } H}{4 k^3}-\frac{3 \sqrt{\pi } H (-\tau_1)^2}{4 k} - \frac{i}{2} \sqrt{\pi } H (-\tau_1)^3 ~,
		\\
		B_2 &= \frac{3 \sqrt{\pi } H e^{-2 i k \tau_1}}{4 k^3}
		- \frac{3 i \sqrt{\pi } H (-\tau_1)	e^{-2 i k \tau_1}}{2 k^2}-\frac{3 \sqrt{\pi } H (-\tau_1)^2 e^{-2 i k \tau_1}}{4 k} ~.
	\end{align}
\end{subequations}
Plugging the above coefficients into the full solution~\eqref{eq:sol2}, we obtain the evolution of the mode function after the transition $\tau_{1}$.
Similarly, we can derive the coefficients $A_3$ and $B_3$ for the last SR phase using the junction method,
\begin{subequations} \label{eq:A3B3}
		\begin{align}
			A_3 &= - {\sqrt{\pi } H e^{-2 i k \tau_{1}} \over 8 k^6 \tau_{2}^6} \Big[ e^{2 i k \tau_{1}} \left(2 k^3 \tau_{1}^3+3 i k^2 \tau_{1}^2+3 i\right) \left(2 k^3 \tau_{2}^3-3 i k^2 \tau_{2}^2-3 i\right) 
			\nn
            \\& \quad\quad\quad -9 e^{2 i k \tau_{2}} (k\tau_{1}-i)^2 (k \tau_{2}+i)^2 \Big] ~,
			\\
			B_3 &= {3 \sqrt{\pi } H e^{-2 i k (\tau_{1}+\tau_{2})} \over 8 k^6 \tau_{2}^6} \Big[ e^{2 i k \tau_{1}} \left(-2 i k^3 \tau_{1}^3+3 k^2 \tau_{1}^2+3\right) (k \tau_{2}-i)^2 
            \nn
			\\& \quad\quad\quad + i e^{2 i k \tau_{2}}  (k \tau_{1}-i)^2 \left(2 k^3 \tau_{2}^3+3 i k^2 \tau_{2}^2+3
			i\right) \Big] ~.
		\end{align}
\end{subequations}

In the following discussion, we aim to derive the appropriate asymptotic expressions of the power spectrum $\mathcal{P}_{\chi}(\tau, k)$ in three distinct $k$ regimes illustrated in Fig.~\ref{fig:k_regimes}. The asymptotic expressions of $\mathcal{P}_{\chi}(\tau, k)$ can be derived by expanding the Hankel functions $H_{d}^{(1,2)}(- k \tau)$ on either subhorizon ($-k\tau \gg 1$) or superhorizon ($-k\tau \ll 1$) scales. In particular, we first establish that a universal conclusion holds for the superhorizon case, applicable for general $d$.

\subsection{Asymptotics on superhorizon scales: finite growing modes}

For a non-integer $d$, the power series of Hankel functions are given by~\cite{olver2010nist},
\begin{subequations} \label{eq:super_series_Hankel}
	\begin{align}
		H_{d}^{(1)}(z)
		&= {i \Gamma(d) \over \pi} \sum_{m=0}^{\infty} {(-1)^{m+1} 2^{d-2m} \over m! (1-d)_{m} } z^{2m-d}
		  + [1 + i \cot(d\pi)] \sum_{m=0}^{\infty} {(-1)^m 2^{-d-2m} \over m!\Gamma(m+d+1)} z^{2m+d}  ~,
		\\
		H_{d}^{(2)}(z)
		&= - {i \Gamma(d) \over \pi} \sum_{m=0}^{\infty} {(-1)^{m+1} 2^{d-2m} \over m! (1-d)_{m} } z^{2m-d}
		+ [1 - i \cot(d\pi)] \sum_{m=0}^{\infty} {(-1)^m 2^{-d-2m} \over m!\Gamma(m+d+1)} z^{2m+d} ~,
	\end{align}
\end{subequations}
which converge rapidly when $z \ll 1$, making them well-suited for the asymptotic analysis of mode functions on superhorizon scales.
Here, $(a)_{m} \equiv a (a+1) \cdots (a+m-1)$ is the Pochhammer symbol. 
Plugging Eq.~\eqref{eq:super_series_Hankel} into Eq.~\eqref{eq:chi_sol_AB}, we derive
\begin{equation} \label{eq:series_chi}
	\begin{aligned}
		\chi_{k}(\tau) &=
		{i (A-B) \Gamma(d) \over \pi} \sum_{m=0}^{\infty} {(-1)^{m+1} 2^{d-2m} \over m! (1-d)_{m} } k^{2m-d} (-\tau)^{2m}
		\\& \quad +
		\l[ A + B + i \cot(d\pi) (A-B) \r]  \times\sum_{m=0}^{\infty} {(-1)^m 2^{-d-2m} \over m!\Gamma(m+d+1)} k^{2m+d} (-\tau)^{2m+2d} ~.
	\end{aligned}
\end{equation}

Examining Eq.~\eqref{eq:super_series_Hankel}, the asymptotics of Hankel functions on superhorizon scales (i.e., $z \rightarrow 0$) can be written concisely as $H_{d}^{(1,2)}(z) \rightarrow (z^{d}, z^{-d}) \times (z^{0}, z^{2}, z^{4}, z^{6}, \cdots)$, where dots denote higher-order terms in $z$, all of which are even powers. Consequently,
\begin{equation} \label{eq:hankel}
	\begin{aligned}
		z^{d} H_{d}^{(1,2)}(z) \rightarrow& (z^{2d}, z^{2d+2}, z^{2d+4}, z^{2d+6}, \cdots) + ( z^{0}, z^{2}, z^{4}, z^{6}, \cdots ) ~,
	\end{aligned}
\end{equation}
which demonstrates that only the first branch, the sequence $(z^{2d}, z^{2d+2}, z^{2d+4}, z^{2d+6}, \cdots)$, can exhibit growth when $d$ is negative. 
{\it Hence, for each such $d$, the set of growing modes is necessarily finite. }
Because the necessary truncation order depends on the specific problem, no universal prescription can be offered. Each case must be evaluated individually.

In the SR phase with $d=3/2$, Eq.~\eqref{eq:hankel} yields $z^{3/2} H_{3/2}^{(1,2)}(z)\rightarrow (z^{3}, z^{5}, z^{7}, z^{9}, \cdots) + ( z^{0}, z^{2}, z^{4}, z^{6}, \cdots )$, and its squared norm is expressed as
\begin{equation}
	\begin{aligned}
		|z^{3/2} H_{3/2}^{(1,2)}(z)|^2 
		&\rightarrow
		z^{0} \{z^{0} z^{0}\} + z^{2} \{z^{0} z^{2}\} + z^{3} \{z^{0} z^{3}\} + z^{4} \{z^{0} z^{4}, z^{2} z^{2}\} 
		\\& \quad + z^{5} \{z^{0} z^{5}, z^{2} z^{3}\} + z^{6} \{z^{0} z^{6}, z^{2} z^{4}, z^{3} z^{3} \} ~,
	\end{aligned}
\end{equation}
where we have dropped irrelevant higher-order terms. 
Each pair of curly brackets indicates how the corresponding power in the expansion is formed from products of terms in $|z^{3/2} H_{3/2}^{(1,2)}(z)|^2$. 
For instance, the product of terms $z^{0}$ and $z^{2}$ in $z^{3/2} H_{3/2}^{(1,2)}(z)$ gives the mode $z^{2}$.
As expected, no growing mode is present in the SR phase.
Similarly, in the USR phase with $d=-3/2$, Eq.~\eqref{eq:hankel} yields
\begin{equation} \label{eq:hankel_expansion}
	\begin{aligned}
		|z^{-3/2} H_{-3/2}^{(1,2)}(z)|^2 
		&\rightarrow
		z^{-6} \{z^{-3} z^{-3}\} + z^{-4} \{z^{-3} z^{-1}\} + z^{-3} \{z^{-3} z^{0}\} + z^{-2} \{z^{-3} z^{1}, z^{-1} z^{-1}\} 
        \\& \quad + z^{-1} \{z^{-3} z^{2}, z^{-1} z^{0}\} + z^{0} \{z^{-3} z^{3}, z^{-1} z^{1}, z^{0} z^{0}\} ~.
	\end{aligned}
\end{equation}
{\it Hence, retaining terms up to $z^{2d + 6} = z^{3}$ in $\chi_{k}$ suffices to capture all growing modes of the USR solution, as well as the complete constant mode.} 
Compared with the SR case, the constant mode now receives two additional contributions, i.e., $z^{-3} z^{3}$ and $z^{-1} z^{1}$. The above facts also highlight the importance of mixing terms when considering the asymptotic behaviors of power spectra.

Based on the preceding analysis, we calculate the superhorizon asymptotics ($-k\tau \ll 1$) of power spectra in both the SR and USR phases, retaining all relevant terms. Applying the power series~\eqref{eq:series_chi}, in contrast to existing literature, we retain the power spectrum during the SR phase up to the order $(-\tau)^6$, and obtain
\begin{equation} \label{eq:sr_super_AB}
	\begin{aligned}
		\lim_{-k\tau \ll 1} \mathcal{P}_{\chi}^{\rm SR}(\tau, k)
		&\simeq
		{|A-B|^2 \over \pi^3}
		+ { 2 i (A^* B - A B^*) \over 3 \pi^3} k^3 (-\tau)^3
		+ { 2 (A^* B + A B^*) \over 9 \pi^3} k^6 (-\tau)^6 ~.
	\end{aligned}
\end{equation}
Similarly, in the USR phase with $d=-3/2$, the power spectrum is obtained as
\begin{equation} \label{eq:usr_super_AB}
	\begin{aligned}
		\lim_{-k\tau \ll 1} \mathcal{P}_{\chi}^{\rm USR}(\tau, k)
		&\simeq
		- {2(A B^* + A^* B) \over 9 \pi^3} k^6
		+ { |A|^2 + |B|^2 + 129 (A^* B + A B^*) \over 2880 \pi^3} k^8 (-\tau)^2
		\\& \quad
		+ {2 i \l(A B^* - A^* B \r) \over 3 \pi^3} k^{3} (-\tau)^{-3}
		+ {|A+B|^2 \over \pi^3} (-\tau)^{-6} ~,
	\end{aligned}
\end{equation}
which includes all growing modes as well as the complete constant mode. And we retain only the leading-order terms that share common coefficients of $A$ and $B$.
It should be noted that the limit notation hereafter denotes asymptotic behavior, not a strict mathematical limit.

Additionally, for the SR phase with $d=3/2$, one can show that the power series in Eq.~\eqref{eq:super_series_Hankel} resums to yield,
\begin{equation} \label{eq:simple_hankel_sr}
	\begin{aligned}
		H_{3/2}^{(1)}(z) &= e^{i z} z^{-3/2} \sqrt{2 \over \pi} (-i - z) ~,
		\quad
		H_{3/2}^{(2)}(z) &= e^{-i z} z^{-3/2} \sqrt{2 \over \pi} (i - z) ~,
	\end{aligned}
\end{equation}
and for the USR phase with $d=-3/2$, Eq.~\eqref{eq:super_series_Hankel} reduces to
\begin{equation} \label{eq:simple_hankel_usr}
	\begin{aligned}
		H_{-3/2}^{(1)}(z) &= i e^{i z} z^{-3/2} \sqrt{2 \over \pi} (i + z) ~,
		\quad
		H_{-3/2}^{(2)}(z) &= -i e^{-i z} z^{-3/2} \sqrt{2 \over \pi} (-i + z) ~.
	\end{aligned}
\end{equation} 
In Appendix~\ref{app:check}, we demonstrate that all the relevant terms (including the constant mode and growing modes) of power spectra~\eqref{eq:sr_super_AB} and \eqref{eq:usr_super_AB} can be derived consistently from the exact expressions, based on the compact forms in Eqs.~\eqref{eq:simple_hankel_sr} and~\eqref{eq:simple_hankel_usr}.

\subsection{Asymptotics on subhorizon scales}

Beginning with the power series given in Eq.~\eqref{eq:super_series_Hankel}, a form suitable for asymptotic analysis on subhorizon scales ($z \gg 1$) can be derived as~\cite{olver2010nist},
\begin{subequations} \label{eq:sub_series_Hankel}
	\begin{align}
		H_{d}^{(1)}(z) &=
		\sqrt{2 \over \pi} e^{i \l(z - {d \pi \over 2} - {\pi \over 4} \r)} z^{-1/2} \l[ 1+ \sum_{n=1}^{\infty} {(1/2 - d)_{n} (1/2 + d)_{n} \over (2 i)^n n!} z^{-n} \r] ~,
		\\
		H_{d}^{(2)}(z) &=
		\sqrt{2 \over \pi} e^{-i \l(z - {d \pi \over 2} - {\pi \over 4} \r)} z^{-1/2} \l[1 + \sum_{n=1}^{\infty} {(1/2 - d)_{n} (1/2 + d)_{n} \over (-2 i)^n n!} z^{-n} \r] ~.
	\end{align}
\end{subequations}
Hence, in the subhorizon limit $-k \tau \gg 1$, we expand the general solution~\eqref{eq:chi_sol_AB} as
\begin{equation} \label{eq:chi_sub_AB}
		\begin{aligned}
			\lim_{-k\tau \gg 1} \chi_{k}(\tau) 
			&\simeq
			\sqrt{2 \over \pi} \l[ A e^{- i k \tau} e^{ -i \l( {d \pi \over 2} + {\pi \over 4} \r)} + B e^{i k \tau} e^{i \l( {d \pi \over 2} + {\pi \over 4} \r)} \r] k^{-1/2} (- \tau)^{d - 1/2}
			\\&
			+ \sqrt{2 \over \pi} \sum_{n=1}^{\infty} {(1/2 - d)_{n} (1/2 + d)_{n} \over (2 i)^n n!} \l[ A e^{- i k \tau} e^{ -i \l( {d \pi \over 2} + {\pi \over 4} \r)}  + (- 1)^{n} B e^{i k \tau} e^{i \l( {d \pi \over 2} + {\pi \over 4} \r)} \r] k^{-n-1/2} (- \tau)^{d -n - 1/2} ~.
		\end{aligned}
\end{equation}
Notably, the series expansion in Eq.~\eqref{eq:chi_sub_AB} truncates at $n=1$ for both SR and USR phases, due to the fact that $(1/2 - d)_{2} (1/2 + d)_{2} = 0$ for $d = \pm 3/2$. Hence, we obtain
\begin{equation}
	\begin{aligned}
		\chi_{k}^{\rm SR}(\tau) 
		&=
		\sqrt{2 \over \pi} \l( - A e^{- i k \tau} - B e^{i k \tau} \r) k^{-1/2} (- \tau)
		+ i \sqrt{2 \over \pi} \l( - A e^{- i k \tau} + B e^{i k \tau} \r) k^{-3/2} ~,
	\end{aligned}
\end{equation}
and
\begin{equation}
	\begin{aligned}
		\chi_{k}^{\rm USR}(\tau) 
		&=
		i \sqrt{2 \over \pi} \l( A e^{- i k \tau} - B e^{i k \tau} \r) k^{-1/2} (- \tau)^{-2}
		+ \sqrt{2 \over \pi} \l( - A e^{- i k \tau} - B e^{i k \tau} \r) k^{-3/2} (-\tau)^{-3} ~,
	\end{aligned}
\end{equation}
respectively. 
Importantly, these two expressions already coincide with the exact Hankel functions given in Eqs.~\eqref{eq:simple_hankel_sr} and~\eqref{eq:simple_hankel_usr}. Therefore, we employ the full expressions for the subhorizon asymptotics and omit the limiting notation here.

In the following discussion, we only focus on the expansion at $n=1$ in Eq.~\eqref{eq:chi_sub_AB}.
The power spectrum is thus calculated as
\begin{equation} \label{eq:sub_AB}
	\begin{aligned}
		\mathcal{P}_{\chi}(\tau, k)&= 
		{1\over\pi^3} k^{2} (-\tau)^{2d - 1} \Big[ X_{\rm non} + i X_{d,-}\cos(2k\tau) - X_{d,+}\sin(2k\tau) \Big]
		\\&\quad
		+ {4 d^2 -1 \over 4 \pi^3} k (-\tau)^{2 d - 2} \Big[ X_{d,+} \cos(2k\tau) + i X_{d,-} \sin(2k\tau) \Big]
		\\&\quad
		+ {1 \over \pi^3} \l( {4 d^2 -1 \over 8} \r)^2 (-\tau)^{2 d - 3} \times\Big[ X_{\rm non} - i X_{d,-}\cos(2k\tau) + X_{d,+}\sin(2k\tau) \Big] ~,
	\end{aligned}
\end{equation}
where $X_{d,\pm} \equiv A^* B e^{i d\pi} \pm A B^* e^{-i d\pi}$ represents the amplitudes of oscillations $\sin(2k\tau)$ and $\cos(2k\tau)$, respectively. While $X_{\rm non} \equiv |A|^2 + |B|^2$ corresponds to the non-oscillating piece of $\mathcal{P}_{\chi}(\tau, k)$.
In the SR phase with $d=3/2$, Eq.~\eqref{eq:sub_AB} gives,
\begin{equation} \label{eq:sr_sub_AB}
	\begin{aligned}
		\mathcal{P}_{\chi}^{\rm SR}(\tau, k)
		&=
		{1\over\pi^3} k^{2} (-\tau)^{2} \Big[ X_{\rm non} + i X_{3/2,-} \cos(2k\tau) - X_{3/2,+}\sin(2k\tau) \Big]
		\\&\quad
		+ {2 \over \pi^3} k (-\tau) \Big[ i X_{3/2,-} \sin(2k\tau) + X_{3/2,+} \cos(2k\tau) \Big]
		\\&\quad
		+ {1 \over \pi^3} \Big[ X_{\rm non} - i X_{3/2,-}\cos(2k\tau) + X_{3/2,+}\sin(2k\tau) \Big] ~,
	\end{aligned}
\end{equation}
where $X_{3/2,\pm} = i (- A^* B \pm A B^*)$.
In the USR phase with $d=-3/2$, Eq.~\eqref{eq:sub_AB} gives,
\begin{equation} \label{eq:usr_sub_AB}
	\begin{aligned}
		\mathcal{P}_{\chi}^{\rm USR}(\tau, k)
		&= 
		{1\over\pi^3} k^{2} (-\tau)^{-4} \Big[ X_{\rm non} + i X_{-3/2,-}\cos(2k\tau) - X_{-3/2,+}\sin(2k\tau) \Big]
		\\&\quad
		+ {2 \over \pi^3} k (-\tau)^{-5} \Big[ X_{-3/2,+} \cos(2k\tau) + i X_{-3/2,-} \sin(2k\tau) \Big]
		\\&\quad
		+ {1 \over \pi^3} (-\tau)^{-6} \Big[ X_{\rm non} - i X_{-3/2,-}\cos(2k\tau) + X_{-3/2,+}\sin(2k\tau) \Big] ~,
	\end{aligned}
\end{equation}
where $X_{-3/2,\pm} = -i(-A^* B \pm A B^*)$.

Based on the superhorizon asymptotics in Eqs.~\eqref{eq:sr_super_AB} and \eqref{eq:usr_super_AB}, the subhorizon asymptotics in Eqs.~\eqref{eq:sr_sub_AB} and \eqref{eq:usr_sub_AB}, and the full expressions for coefficients in Eqs.~\eqref{eq:A2B2} and \eqref{eq:A3B3}, our next step is to expand the coefficients appearing in these results and retain leading-order contributions.
This will allow us to give a clear analytical explanation of the characteristic features of the final power spectrum in distinct $k$ regimes.

\section{Time Evolutions of Power Spectra}\label{sec:asymptotic}

\subsection{Case 1}

We first examine Case 1 illustrated in Fig.~\ref{fig:k_regimes}, in which the mode exits horizon during the first SR phase, such that $-k \tau_{2} < - k\tau_{1} \ll 1$ (or equivalently, $k \ll 1/(-\tau_{1})$), where $\tau_{1,2}$ denote the first and second transition times, respectively.
Our goal is to determine appropriate superhorizon asymptotics of the power spectra~\eqref{eq:sr_super_AB} and~\eqref{eq:usr_super_AB} during the intermediate USR phase and the last SR phase. In the first SR phase, one can directly apply the results given in Eqs.~\eqref{eq:sr1_full}, \eqref{eq:sr1_sub} and \eqref{eq:sr1_super}.

\subsubsection{USR}

\begin{figure}[h]
	\centering
	\includegraphics[width=0.35\textheight]{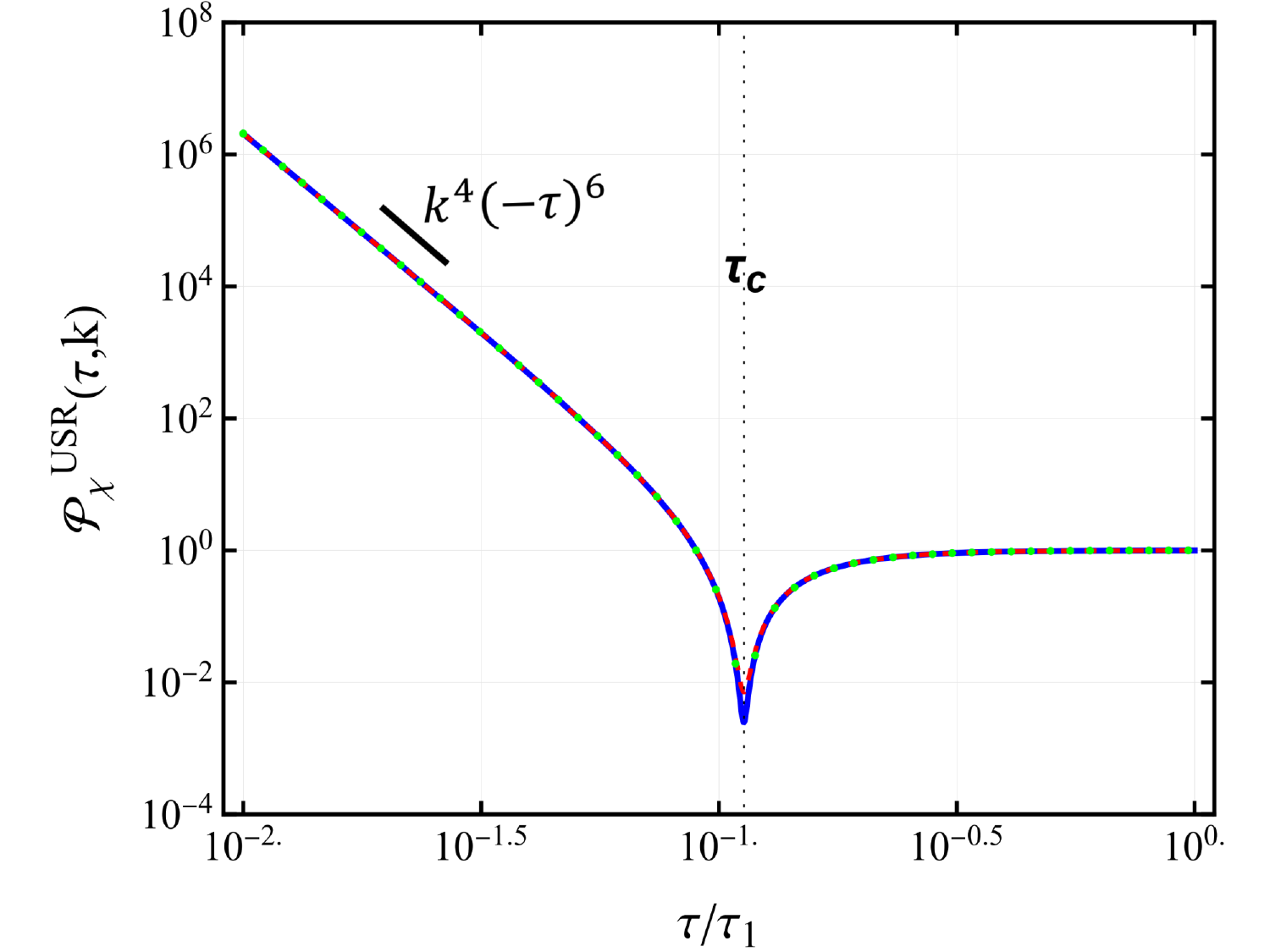}
	\caption{
		The comparison among the exact solution~\eqref{eq:chi_sol_AB} (the blue solid curve) with coefficients~\eqref{eq:A2B2}, the asymptotic solution~\eqref{case1:usr_super} (the red dashed curve, with the parameter choices: $k = 0.06$ and $\tau_1 = -1$), and the numerical result (the green dots) of $\mathcal{P}_{\chi}^{\rm USR}(\tau, k)$ based on the reconstructed SR-USR-SR inflationary potential~\eqref{eq:final_potential}. The value of $\mathcal{P}_{\chi}^{\rm USR}(\tau, k)$ is normalized by the exact value of $\mathcal{P}_{\chi}^{\rm USR}(\tau_{1}, k)$ based on Eq.~\eqref{eq:chi_sol_AB}.
		The black dotted vertical line refers to the dip time $\tau_c$ given by Eq.~\eqref{case1:tauc}. 
		The observed dip structure around $\tau_{c}$ emerges from the interplay between the positive $(-\tau)^{-6}$ term and negative $(-\tau)^{-3}$ term in Eq.~\eqref{case1:usr_super}.
	}
	\label{fig:usr_chi_general}
\end{figure}

For the USR phase, we evaluate the coefficients in Eq.~\eqref{eq:usr_super_AB} in the superhorizon limit imposed at the first transition, $- k \tau_{1} \ll 1$. 
Following three Rules in Introduction and inserting the exact solutions in Eq.~\eqref{eq:A2B2} into Eq.~\eqref{eq:usr_super_AB}, we obtain the following leading-order contributions,
\begin{subequations}
	\begin{align}
		\lim_{- k \tau_1 \ll 1} \l( A_{2} B_{2}^* + A_{2}^* B_{2} \r)
		&\simeq - {9 \pi H^2 \over 8} k^{-6} - {9 \pi H^2 \over 4} k^{-4} (-\tau_{1})^{2} ~,
		\\
		\lim_{- k \tau_1 \ll 1} i \l( A_{2} B_{2}^* - A_{2}^* B_{2} \r)
		&\simeq - {3 \pi H^2 \over 10} k^{-1} (-\tau_{1})^{5} ~,
		\\
		\lim_{- k \tau_1 \ll 1} \l(|A_{2}|^2 + |B_{2}|^2 + 129 (A_{2}^* B_{2} + A_{2} B_{2}^*) \r)
		&\simeq
		- 144 \pi H^2 k^{-6} ~,
		\\
		\lim_{- k \tau_1 \ll 1} |A_{2}+B_{2}|^2 
		&\simeq {\pi H^2 \over25} (- \tau_{1})^{10} k^{4} ~.
	\end{align}
\end{subequations}
and substituting these expressions into Eq.~\eqref{eq:usr_super_AB} yields, 
\begin{equation} \label{case1:usr_super}
		\begin{aligned}
			\lim_{-k\tau \ll 1} \mathcal{P}_{\chi}^{\rm USR}(\tau, k)
			\simeq
			\mathcal{P}_{\chi,{\rm CMB}} \l[ 1 + 2(-\tau_{1})^{2} k^{2} - {1\over 5} k^{2} (-\tau)^2 
			- {4(-\tau_{1})^5 \over 5}  k^{2} (-\tau)^{-3} + {4 (-\tau_{1})^{10} \over 25} k^{4} (-\tau)^{-6} \r] ~.
		\end{aligned}
\end{equation}
Figure~\ref{fig:usr_chi_general} shows the comparison between this asymptotic solution (the red dashed curve) and the exact solution~\eqref{eq:chi_sol_AB} with the coefficients~\eqref{eq:A2B2} (the blue solid curve).
We normalize the numerical results by setting $\tau_{1} = - 1$ throughout the paper, while the symbolic analyses retain the parameter $\tau_{1}$ to ensure generality. Also, we normalize the value $\mathcal{P}_{\chi}^{\rm USR}(\tau, k)$ by the exact value of $\mathcal{P}_{\chi}^{\rm USR}(\tau_{1}, k)$ in Fig.~\ref{fig:usr_chi_general}.

Following Rule 1\&3 in Introduction, one can straightforwardly identify the dominant terms in Eq.~\eqref{case1:usr_super} at the transition time $\tau_{1}$:
\begin{equation} \label{case1:rule1}
	\begin{aligned}
		{H^2 \over 4\pi^2}
		> {H^2 \over 2\pi^2} (-\tau_{1})^{2} k^{2}
		> {H^2 (-\tau_{1})^5 \over 5 \pi^2} k^{2} (-\tau)^{-3}
		> {H^2 \over 20 \pi^2} k^{2} (-\tau)^2
		> {H^2 (-\tau_{1})^{10} \over 25\pi^2} k^{4} (-\tau)^{-6} ~.
	\end{aligned}
\end{equation}
In contrast to the usual expansion in literature~\cite{Carrilho:2019oqg}, we need to retain the next-to-leading-order $k$-dependent term, ${H^2 \over 2\pi^2} (-\tau_{1})^{2} k^{2}$, in the constant mode of Eq.~\eqref{case1:usr_super}, as well as the decaying $(-\tau)^2$ mode. 
At the transition time $\tau_{1}$, the term ${H^2 \over 2\pi^2} (-\tau_{1})^{2} k^{2}$ dominates all other growing modes in Eq.~\eqref{case1:usr_super}, and later produces a finite dip in the time evolution of $\mathcal{P}_{\chi}^{\rm USR}(\tau, k)$,
\footnote{The decaying $(-\tau)^{2}$ term is negligible at the dip time $\tau_{c}$ defined by Eq.~\eqref{case1:tauc}.}
see Eq.~\eqref{eq:case1_dip_value} and Fig.~\ref{fig:usr_chi_general}.
Moreover, Eq.~\eqref{case1:usr_super} at $\tau_{1}$ gives
\begin{equation} \label{case1:usr_tau1}
	\mathcal{P}_{\chi}^{\rm USR}(\tau_{1}, k)
	\simeq
	\mathcal{P}_{\chi,{\rm CMB}} \l[ 1 + (-k\tau_{1})^2 \r] ~,
\end{equation}
which agrees with the exact SR solution~\eqref{eq:sr1_full}, as required by the junction condition~\eqref{eq:junction_condition}. This consistency would be lost were the decaying $(-\tau)^2$ term in Eq.~\eqref{case1:usr_super} omitted, because this term also dominates the growing $(-\tau)^{-6}$ mode at the transition, refer to Eq.~\eqref{case1:rule1}.

The asymptotic power spectrum~\eqref{case1:usr_super} shows that the dominant late-time growth is carried by the growing $(-\tau)^{-6}$ mode with a $k^4$-dependence, see Fig.~\ref{fig:usr_chi_general}. Crucially, the negative $(-\tau)^{-3}$ term dominates over the $(-\tau)^{-6}$ term initially, refer to Eq.~\eqref{case1:rule1}. Then after the transition, the $(-\tau)^{-3}$ term first cancels the constant mode $\mathcal{P}_{\chi,{\rm CMB}} \l[ 1 + 2(-\tau_{1})^{2} k^{2}\r]$, until the difference between the $(-\tau)^{-6}$ and $(-\tau)^{-3}$ terms attains its extremum, leading to the dip structure as shown in Fig.~\ref{fig:usr_chi_general}. Then, we can calculate the corresponding dip time as
\begin{equation} \label{case1:tauc}
	\tau_c = -\l( {2 \over 5} \r)^{1/3} k^{2/3} (-\tau_{1})^{5/3} ~,
\end{equation}
indicated by the black dotted vertical lines in Fig.~\ref{fig:usr_chi_general}, and in excellent agreement with the exact result (the blue solid curve). 
Then, we calculate the dip value as
\begin{equation} \label{eq:case1_dip_value}
	\mathcal{P}_{\chi}^{\rm USR}(\tau_{c}, k) \simeq 2\mathcal{P}_{\chi,{\rm CMB}} (- k \tau_{1})^2 ~,
\end{equation}
which is finite. Note that $-\tau_c < -\tau_1$ holds for $- k \tau_{1} < 1$, meaning the dip of the time evolution of USR mode always exists after the transition time $\tau_{1}$ (as far as the USR phase lasts sufficiently long duration, refer to Fig.~\ref{fig:case1}) in Case 1.

\subsubsection{SR2}\label{subsec:case1_sr2}

\begin{figure*}[h]
	\centering
	\includegraphics[width=0.3\textheight]{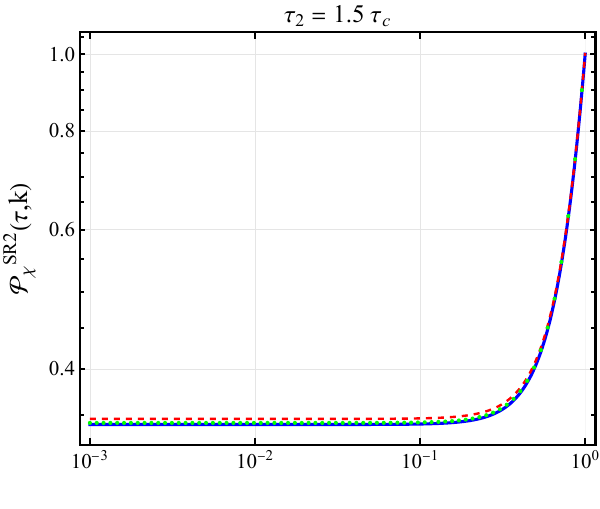}
	\includegraphics[width=0.3\textheight]{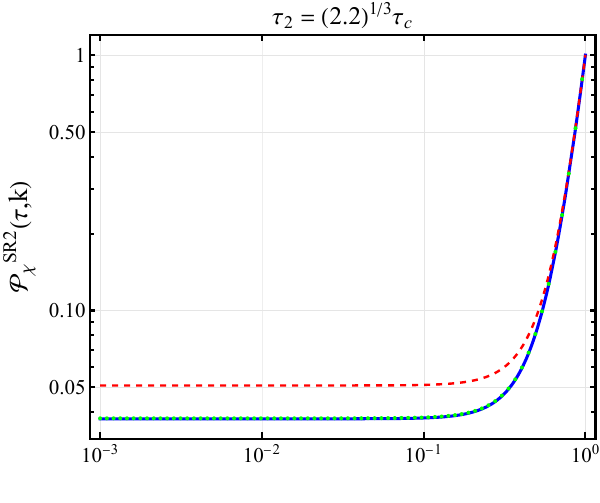}
	\includegraphics[width=0.3\textheight]{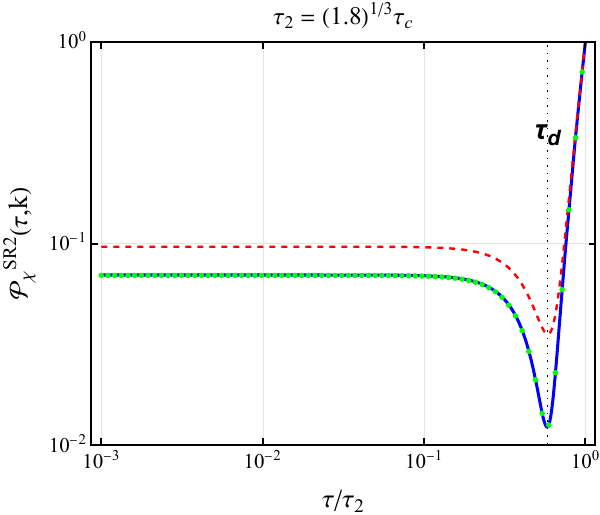}
	\includegraphics[width=0.3\textheight]{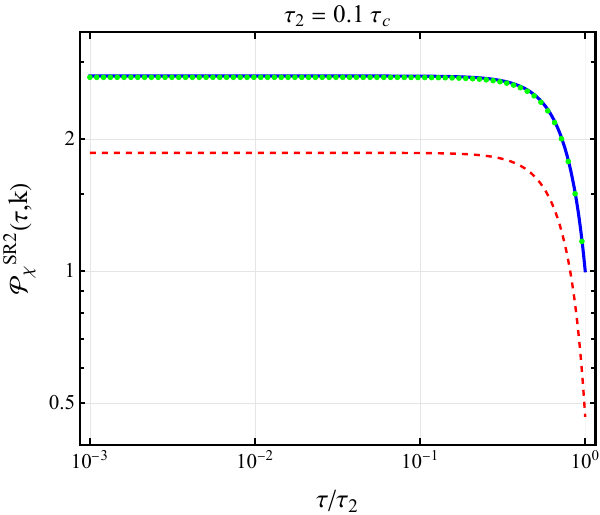}
	\caption{
		The comparisons among the exact solution~\eqref{eq:chi_sol_AB} (the blue solid curves) with the coefficients~\eqref{eq:A3B3}, the asymptotic power spectrum~\eqref{case1:sr2_super} (the red dashed curves) for $k = 0.06$ and $\tau_1 = -1$ with $\tau_{2}$ scaled as $1.5 \tau_{c}$, $(2.2)^{1/3} \tau_{c}$, $(1.8)^{1/3} \tau_{c}$, $0.1 \tau_{c}$, and the numerical result (the green dots) of power spectrum based on the reconstructed SR-USR-SR inflationary potential~\eqref{eq:final_potential}.
		All the results are normalized by the exact value of $\mathcal{P}_{\chi}^{\rm SR2}(\tau_{2}, k)$ based on Eq.~\eqref{eq:chi_sol_AB}.
		The dip time $\tau_{d}$ is given by Eq.~\eqref{case1:taud}, which only exists when $-\tau_{2} < -2^{1/3} \tau_{c}$.
	}
	\label{fig:sr2_super}
\end{figure*}

At the second transition time $\tau_{2}$, we require the superhorizon asymptotics of the coefficients in Eq.~\eqref{eq:A3B3} under $-k \tau_2 < - k \tau_1 \ll 1$. 
Treating it as a two-parameter expansion, we introduce two small bookkeeping parameters, $x \equiv - k \tau_1$ and $y \equiv -k \tau_2$, where $y < x \ll 1$. Then, we rewrite $A_3$ and $B_3$ in Eq.~\eqref{eq:A3B3} as
\begin{subequations} \label{eq:A3B3xy}
		\begin{align}
			A_3 &= - {\sqrt{\pi } H e^{2 i x} \over 8 y^6} \l[ e^{- 2 i x} \left(-2 x^3 + 3 i x^2+3 i\right) \left(- 2 y^3-3 i y^2-3 i\right)-9 e^{-2 i y} (-x-i)^2 (-y+i)^2 \r] ~,
			\\
			B_3 &= {3 \sqrt{\pi } H e^{2 i (x + y)} \over 8 y^6} \Big[ e^{- 2 i x} \left(2 i x^3+3 x^2+3\right) (-y-i)^2 + i e^{-2 i y}  (-x-i)^2 \left(-2 y^3+3 i y^2+3i\right) \Big] ~.
		\end{align}
\end{subequations}
Expanding to the leading order in $y < x \ll 1$ gives
\begin{subequations}
		\begin{align}
			\lim_{- k \tau_2 < - k \tau_1 \ll 1} |A_{3}-B_{3}|^2 
			&\simeq
			\pi H^2 \l[ {1 \over 4} + {1\over2} k^2(-\tau_{1})^2 - {2 \over 5} {(-\tau_{1})^{5} \over (-\tau_{2})^{3}} k^2 + {4 \over 25} {(-\tau_{1})^{10} \over (-\tau_{2})^{6}} k^4 \r] ~,
			\\
			\lim_{- k \tau_2 < - k \tau_1 < 1} i (A_{3}^* B_{3} - A_{3} B_{3}^* )
			&\simeq
			\pi H^2 \l[ - {6  \over 25} {(-\tau_{1})^{10} \over (-\tau_{2})^{9} } k + {3 \over 10} { (-\tau_{1})^5  \over (-\tau_{2})^6} k^{-1} \r] ~,
			\\
			\lim_{- k \tau_2 < - k \tau_1 < 1} (A_{3}^* B_{3} + A_{3} B_{3}^*)
			&\simeq
			{9 \pi  H^2 \over 50} {(-\tau_{1})^{10} \over (-\tau_{2})^{12} } k^{-2} ~.
		\end{align}
\end{subequations}
Inserting these expressions into Eq.~\eqref{eq:sr_super_AB} yields
\begin{equation} \label{case1:sr2_super}
		\begin{aligned}
			\lim_{-k\tau \ll 1} \mathcal{P}_{\chi}^{\rm SR2}(\tau, k)
			\simeq 
			\mathcal{P}_{\chi,{\rm CMB}} &\Bigg\{ \l[ 1 + 2 k^2(-\tau_{1})^2 - {8 \over 5} {(-\tau_{1})^{5} \over (-\tau_{2})^{3}} k^2 + {16 \over 25} {(-\tau_{1})^{10} \over (-\tau_{2})^{6}} k^4 \r]
			\\&
			+ \l[ - {16 \over 25} {(-\tau_{1})^{10} \over (-\tau_{2})^{9} } k^{4} + {4 \over 5} { (-\tau_{1})^5  \over (-\tau_{2})^6} k^{2} \r] (-\tau)^3
			+ {4(-\tau_{1})^{10} \over 25(-\tau_{2})^{12} } k^{4} (-\tau)^6 \Bigg\} ~.
		\end{aligned}
\end{equation}
In contrast to the USR asymptotic solution~\eqref{case1:usr_super}, we keep more terms in Eq.~\eqref{case1:sr2_super}, because the unconstrained ratio $\tau_{2}/\tau_{1}$ makes it impossible to identify dominant terms at the second transition time $\tau_{2}$. Crucially, the sign of $(-\tau)^3$ term in Eq.~\eqref{case1:sr2_super} is determined by the expressions inside the square bracket: it is negative for $0 \lesssim - \tau_{2} < -2^{1/3} \tau_{c}$ and positive for $-2^{1/3} \tau_{c} \lesssim - \tau_{2} < -\tau_{1}$. It is this  sign flip that governs the qualitative shape of $\mathcal{P}_{\chi}^{\rm SR2}(\tau, k)$, refer to Fig.~\ref{fig:sr2_super}.

Before examining the ratio $\tau_{2}/\tau_{1}$, we first calculate the power spectrum at the transition time $\tau_2$ based on Eq.~\eqref{case1:sr2_super}, and obtain
\begin{equation} \label{case1:sr2_super_tau2}
	\begin{aligned}
		\mathcal{P}_{\chi}^{\rm SR2}(\tau_{2}, k)
		\simeq
		\mathcal{P}_{\chi,{\rm CMB}}\times \l[ 1 + 2 k^2(-\tau_{1})^2 
		- {4 \over 5} {(-\tau_{1})^{5} \over (-\tau_{2})^{3}} k^2
		+ {4 \over 25} {(-\tau_{1})^{10} \over (-\tau_{2})^{6} } k^{4} \r] ~,
	\end{aligned}
\end{equation}
which is approximated to the calculation from Eq.~\eqref{case1:usr_super} at $\tau_2$ by discarding the negligible term $H^2 k^2 (-\tau_{2})^2/(20\pi^2)$ arising from the $(-\tau)^{2}$ term in Eq.~\eqref{case1:usr_super}.

Now, we examine the the asymptotic evolution of $\mathcal{P}_{\chi}^{\rm SR2}(\tau, k)$ given by Eq.~\eqref{case1:sr2_super} in terms of the varied ratio $\tau_{2}/\tau_{1}$ for given $\tau_{1}$ and $k$.

\begin{figure*}[htbp]
	\centering
	\includegraphics[width=0.3\textheight]{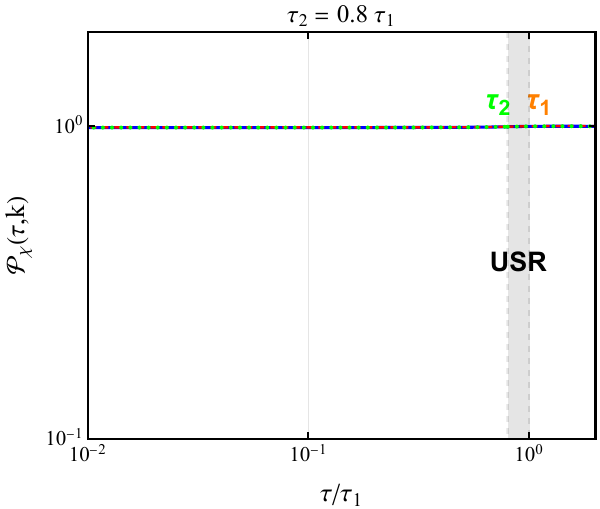}
	\includegraphics[width=0.3\textheight]{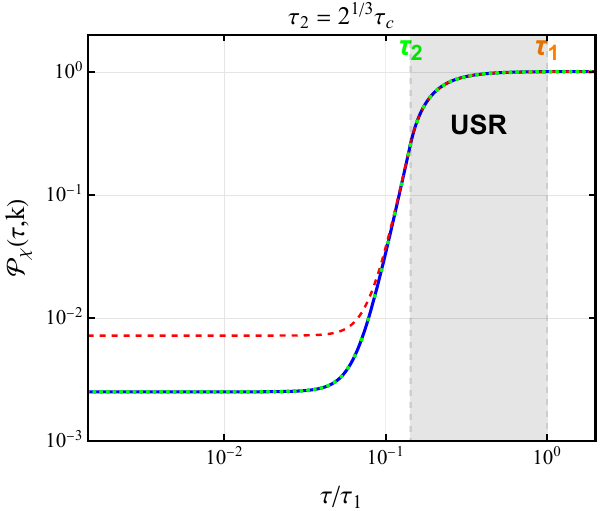}
	\includegraphics[width=0.3\textheight]{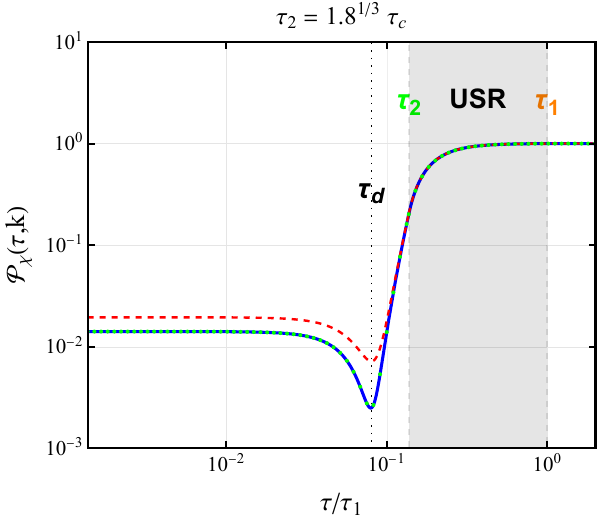}
	\includegraphics[width=0.3\textheight]{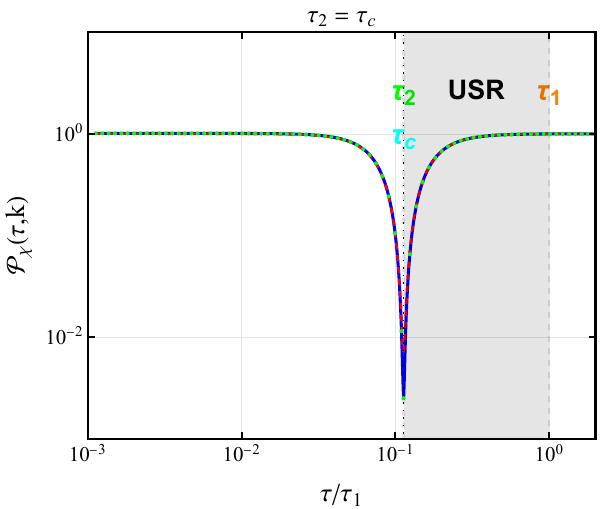}
	\includegraphics[width=0.3\textheight]{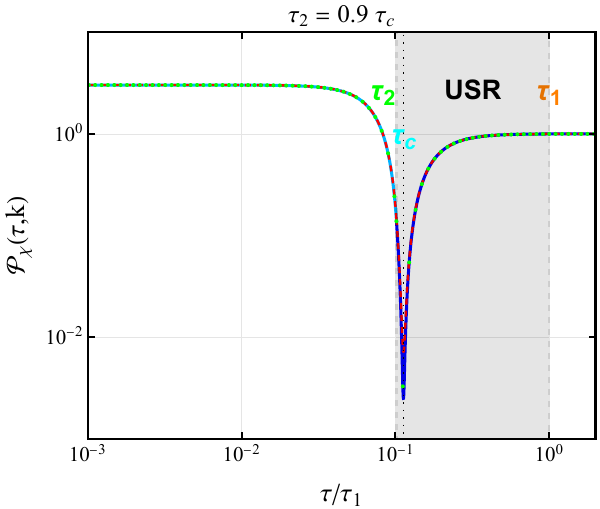}
	\includegraphics[width=0.3\textheight]{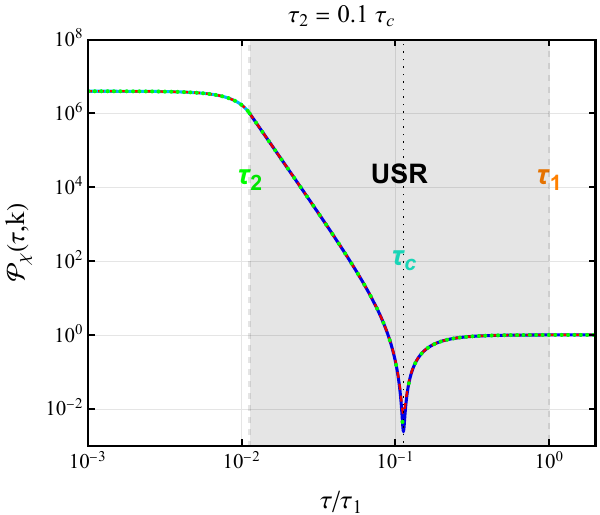}
	\caption{
		Time evolutions of the power spectrum $\mathcal{P}_{\chi}(\tau, k)$ across SR-USR-SR transitions, corresponding to Case 1 of Fig.~\ref{fig:k_regimes} for $\tau_{2} = 0.8\tau_{1}, 2^{1/3}\tau_{c}, 1.8^{1/3}\tau_{c}, \tau_{c}, 0.9\tau_{c}, 0.1\tau_{c}$, respectively. 
		The blue solid curves denote the exact solutions of power spectra based on Eqs.~\eqref{eq:chi_sol_AB}, \eqref{eq:A2B2} and \eqref{eq:A3B3}, while the red dashed curves represent the asymptotic solutions given by Eqs.~\eqref{eq:sr1_full}, \eqref{case1:usr_super} and \eqref{case1:sr2_super}. 
		The green dots refers to the numerical results of $\mathcal{P}_{\chi}(\tau, k)$ based on the reconstructed SR-USR-SR inflationary potential~\eqref{eq:final_potential}.
		All results are normalized by the exact value of $\mathcal{P}_{\chi}^{\rm USR}(\tau_{1}, k)$ based on Eq.~\eqref{eq:chi_sol_AB}.
		The dip times $\tau_c$ and $\tau_{d}$ are defined in Eqs.~\eqref{case1:tauc} and \eqref{case1:taud}, respectively.
	}
	\label{fig:case1}
\end{figure*}

1. $- 2^{2/3} \tau_{c} < - \tau_{2} < -\tau_{1}$,
\footnote{According to Eq.~\eqref{case1:tauc}, it is easy to check that $- 2^{2/3} \tau_{c} < -\tau_{1}$ for $-k \tau_{1} \ll 1$.}
the constant mode dominates the asymptotic behavior of Eq.~\eqref{case1:sr2_super} (expressed for simplicity as $(-\tau)^{0} > (-\tau)^{3} > (-\tau)^{6}$) and the $(-\tau)^3$ term is positive, which corresponds to a quite short USR phase (see the top-left panel of Fig.~\ref{fig:case1}), which leaves the mode function almost unchanged. The power spectrum at the end of inflation can be calculated either by utilizing the constant mode from Eq.~\eqref{case1:usr_super}, or by evaluating the USR solution at $\tau_{2}$ under the approximation $\tau_{2} \simeq \tau_{1}$, both approaches yield identical results. We obtain $\mathcal{P}_{\chi}^{\rm SR2}(\tau_{\rm end}, k) \simeq H^2 / (4\pi^2) = \mathcal{P}_{\chi,{\rm CMB}}$ as expected.

2. $- 3^{1/3} \tau_{c} < - \tau_{2} < - 2^{2/3} \tau_{c}$, we have $(-\tau)^{3} > (-\tau)^{0} > (-\tau)^{6}$ at $\tau_{2}$, and the $(-\tau)^3$ term is positive. 
Hence, the power spectrum $\mathcal{P}_{\chi}^{\rm SR2}(\tau, k)$ first falls off as $(-\tau)^3$ and then settles onto the constant plateau given by Eq.~\eqref{case1:sr2_super_end}. The power spectrum~\eqref{case1:sr2_super} is shown by the red dashed curve in the top-left panel of Fig.~\ref{fig:sr2_super}, along with the exact solution~\eqref{eq:chi_sol_AB} with the coefficients~\eqref{eq:A3B3} represented by the blue solid curve.
All results in Fig.~\ref{fig:sr2_super} are normalized by the exact value of $\mathcal{P}_{\chi}^{\rm SR2}(\tau_{2}, k)$ based on Eq.~\eqref{eq:chi_sol_AB}.

3. $- \l( {5 \over 2} \r)^{1/3} \tau_{c} < - \tau_{2} < - 3^{1/3} \tau_{c}$, we have $(-\tau)^{3} > (-\tau)^{6} > (-\tau)^{0}$ at $\tau_{2}$, and the $(-\tau)^3$ term is positive. The overall evolution is similar to the previous case.

4. $-2^{1/3} \tau_{c} < - \tau_{2} < -\l( {5 \over 2} \r)^{1/3} \tau_{c}$, we have $(-\tau)^{6} > (-\tau)^{3} > (-\tau)^{0}$ at $\tau_{2}$, and the $(-\tau)^3$ term remains positive. 
The power spectrum $\mathcal{P}_{\chi}^{\rm SR2}(\tau, k)$ first decays as $(-\tau)^6$, then switches to a $(-\tau)^3$ fall-off before finally settling at the constant plateau given by Eq.~\eqref{case1:sr2_super_end}. This evolution is illustrated in the top-right panel of Fig.~\ref{fig:sr2_super}.

5. $- \tau_{c} < - \tau_{2} < -2^{1/3} \tau_{c}$, we have $(-\tau)^{6} > (-\tau)^{0} > (-\tau)^{3}$ at $\tau_{2}$, and the $(-\tau)^3$ term turns negative. 
This case is similar to that is shown in Fig.~\ref{fig:usr_chi_general}. 
The power spectrum $\mathcal{P}_{\chi}^{\rm SR2}(\tau, k)$ first falls as $(-\tau)^6$, and is simultaneously driven lower by the negative $(-\tau)^3$ term.
The dip appears when the sum of the $(-\tau)^6$ and $(-\tau)^3$ terms in Eq.~\eqref{case1:sr2_super} reaches its minimum (as shown by the black dotted line in the bottom-left panel of Fig.~\ref{fig:sr2_super}), and we thus calculate the dip time as
\begin{equation} \label{case1:taud}
	\tau_{d} = 2^{-1/3} \tau_{2} \l[ 4 - 5 k^{-2} {(-\tau_{2})^3 \over (-\tau_{1})^5} \r]^{1/3} ~.
\end{equation}
The corresponding dip value is calculated as
\begin{equation}
	\mathcal{P}_{\chi}^{\rm SR2}(\tau_{d}, k) \simeq 2\mathcal{P}_{\chi,{\rm CMB}} (-k \tau_{1})^2 ~,
\end{equation}
which is the same with Eq.~\eqref{eq:case1_dip_value} by accident.
Note that $\tau_d$ only exists when $-\tau_{2} < -2^{1/3} \tau_{c}$, and $-\tau_d < -\tau_{c} < -\tau_{2}$ always holds for $-\tau_{c} < -\tau_{2} < - 2^{1/3} \tau_{c}$ based on the expression~\eqref{case1:taud}, which is clearly shown in the middle-left panel of Fig.~\ref{fig:case1}.
Ultimately, $\mathcal{P}_{\chi}^{\rm SR2}(\tau, k)$ settles to the constant plateau given by Eq.~\eqref{case1:sr2_super_end}. This evolution is shown in the bottom-left panel of Fig.~\ref{fig:sr2_super}. 
Moreover, we find 
\begin{equation}
	\tau_d = \tau_c
	~~\text{when}~~
	\tau_2 = \tau_c ~,
\end{equation}
which means that if the second transition is chosen as $\tau_2 = \tau_c$, we obtain
\begin{equation}
	\tau_2 = \tau_c = \tau_d
	\quad\quad\text{and}\quad\quad
	\mathcal{P}_{\chi}^{\rm SR2}(\tau_{\rm end}, k) = \mathcal{P}_{\chi}^{\rm USR}(\tau_{1}, k) ~,
\end{equation}
as shown in the middle-right panel of Fig.~\ref{fig:case1}. The presence of such a USR phase does not lead to any change of the power spectrum at the end of inflation, corresponding to a typical scale $\sqrt{2}k_{\rm dip}$, as we will discuss in Eq.~\eqref{case1:k_class} and Fig.~\ref{fig:case1k}.

6. $0 < -\tau_{2} < - \tau_{c}$, we have $(-\tau)^{0} > (-\tau)^{6} > (-\tau)^{3}$ at $\tau_{2}$, and $\tau_d$ does not exist anymore, which means the damping effects arising from the negative $(-\tau)^3$ term becomes weaker, the power spectrum $\mathcal{P}_{\chi}^{\rm SR2}(\tau, k)$ grows slightly after the transition $\tau_{2}$, and quickly settles to the constant value given by Eq.~\eqref{case1:sr2_super_end}, as shown in the bottom-right panel of Fig.~\ref{fig:sr2_super}. Importantly, only this regime displays the enhancement of the power spectrum, refer to the bottom-left and -right panels of Fig.~\ref{fig:case1}.

At the end of inflation $\tau_{\rm end}$, the constant mode must dominate, giving
\begin{equation} \label{case1:sr2_super_end}
	\begin{aligned}
		\mathcal{P}_{\chi}^{\rm SR2}(\tau_{\rm end}, k)
		\simeq 
		\mathcal{P}_{\chi,{\rm CMB}}\times \l[ 1 + 2 k^2(-\tau_{1})^2 
		- {8 \over 5} {(-\tau_{1})^{5} \over (-\tau_{2})^{3}} k^2 
		+ {16 \over 25} {(-\tau_{1})^{10} \over (-\tau_{2})^{6}} k^4 \r] ~,
	\end{aligned}
\end{equation}
which encodes the duration of the USR phase. 
Equating Eqs.~\eqref{case1:usr_tau1} and \eqref{case1:sr2_super_end} yields the necessary condition for the enhancement of the power spectrum with respect to the vanilla SR inflation, we derive the equal time as
\begin{equation} \label{eq:tau2eq}
	\tau_{2}^{\rm eq}
	\simeq - \l( {2\over5} \r)^{1/3} k^{2/3} (- \tau_{1})^{5/3}
	= \tau_c ~.
\end{equation}
{\it The enhancement of $\mathcal{P}_{\chi}^{\rm SR2}(\tau_{\rm end}, k)$ therefore requires the second transition to occur after the USR dip, i.e., $- \tau_2 < - \tau_c$.}
This is natural: because the power spectrum~\eqref{case1:usr_super} of USR phase begins to grow only after the dip time $\tau_{c}$ (c.f., Fig.~\ref{fig:usr_chi_general}), and no growing mode persists after the USR phase, as shown in Eq.~\eqref{case1:sr2_super}.

In summary, the asymptotic solutions of $\mathcal{P}_{\chi}(\tau, k)$ for $k \ll 1/(-\tau_{1})$ are provided in Eqs.~\eqref{eq:sr1_full}, \eqref{case1:usr_super} and \eqref{case1:sr2_super}. 
These asymptotic results coincide with the exact solutions (given in Eqs.~\eqref{eq:chi_sol_AB}, \eqref{eq:A2B2} and \eqref{eq:A3B3}), as demonstrated in Fig.~\ref{fig:case1} by the red dashed and blue solid curves, respectively.
The critical time $\tau_c$, defined in Eq.~\eqref{case1:tauc}, emerges from the interplay between the positive growing $(-\tau)^{-6}$ mode and negative growing $(-\tau)^{-3}$ mode of the USR power spectrum~\eqref{case1:usr_super}. 
It determines the second transition time $\tau_2$ that separates parameter regions, in which the final power spectrum $\mathcal{P}_{\chi}^{\rm SR2}(\tau_{\rm end}, k)$ is enhanced or suppressed relative to the exact value of $\mathcal{P}_{\chi}^{\rm SR1}(\tau_{1}, k)$ (equivalently $\mathcal{P}_{\chi}^{\rm USR}(\tau_{1}, k)$):
\begin{equation} \label{case1:class}
	\begin{aligned}
		-\tau_c < &-\tau_2 < -2^{2/3}\tau_{c} ~, && \text{suppression};
		\\
		-2^{2/3}\tau_{c} \ll &-\tau_2 ~\text{and}~ -\tau_2 = -\tau_c ~, && \text{plateau};
		\\
		0 < &-\tau_2 < -\tau_c ~, && \text{enhancement}.
	\end{aligned}
\end{equation}
In other words, the qualitative behaviors of time evolution of the power spectrum across SR-USR-SR transitions is determined by $\tau_{1,2}$ and $k$.

\subsubsection{The dip structure of the final power spectrum}

\begin{figure}[h]
	\centering
	\includegraphics[width=0.35\textheight]{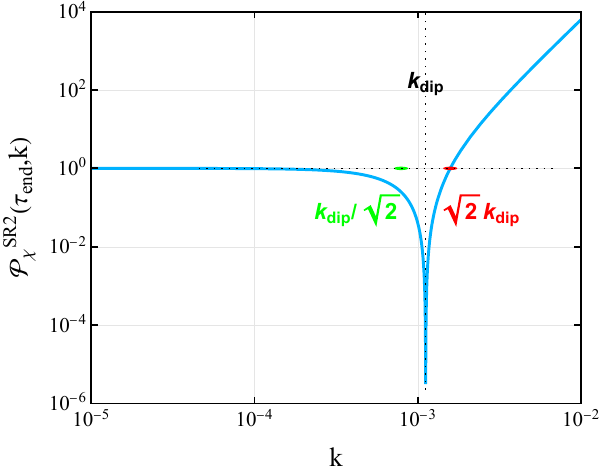}
	\caption{
		The final power spectrum $\mathcal{P}_{\chi}^{\rm SR2}(\tau_{\rm end}, k)$ in Eq.~\eqref{case1:sr2_super_end} with $\tau_{2} = 10^{-2} \tau_{1} = - 0.01$, which features a dip at $k_{\rm dip}$ expressed in Eq.~\eqref{case1:sr2_kdip} or equivalently Eq.~\eqref{case1:sr2_kdip_2}.
		The spectrum is nearly scale-invariant for $k \ll k_{\rm dip}/\sqrt{2}$, and transitions to a $k^4$ growth for $k>\sqrt{2}k_{\rm dip}$.
		The value of $\mathcal{P}_{\chi}^{\rm SR2}(\tau_{\rm end}, k)$ is normalized by $\mathcal{P}_{\chi}^{\rm USR}(\tau_{1}, k)$ based on Eq.~\eqref{eq:chi_sol_AB}.
	}
	\label{fig:case1k}
\end{figure}

Notably, while the preceding analysis is carried out for a fixed $k$, the variation of final power spectrum~\eqref{case1:sr2_super_end} with respect to $k$ can be inferred directly from the conclusion~\eqref{case1:class}.
This is because $\tau_c$ depends on $k$ (c.f., Eq.~\eqref{case1:tauc}), while $\tau_{1,2}$ are $k$-independent model parameters that accompany with $k$ in the combination $-k\tau_{1,2}$. 
{\it Consequently, given transition times $\tau_{1,2}$ map to different dynamical regimes depending on the value of $k$. }
The classification~\eqref{case1:class} is therefore equivalent to
\begin{equation} \label{case1:k_class}
	\begin{aligned}
		{1\over\sqrt{2}}k_{\rm dip} < &k < \sqrt{2} k_{\rm dip} ~, && \text{suppression}~,
		\\
		k \ll {1\over\sqrt{2}}k_{\rm dip} ~\text{and}&~ k = \sqrt{2} k_{\rm dip} ~, && \text{plateau}~,
		\\
		\sqrt{2} k_{\rm dip} < &k < 1/(-\tau_{1}) ~, && \text{enhancement}~.
	\end{aligned}
\end{equation}
As illustrated in Fig.~\ref{fig:case1k}, the final power spectrum $\mathcal{P}_{\chi}^{\rm SR2}(\tau_{\rm end}, k)$ in Eq.~\eqref{case1:sr2_super_end} exhibits a nearly scale-invariant in the small $k$-regime, and transitions to a $k^4$-growth in the relative large $k$-regime. That is, the term ${16 \over 25} {(-\tau_{1})^{10} \over (-\tau_{2})^{6}} k^4$ in Eq.~\eqref{case1:sr2_super_end} dominates in the large $k$-regime.

The dip scale $k_{\rm dip}$ in Eq.~\eqref{case1:k_class}, is obtained from the equation ${\partial \over \partial k} \l[ \mathcal{P}_{\chi}^{\rm SR2}(\tau_{\rm end}, k) \r]\Big|_{k=k_{\rm dip}}= 0$, such that
\begin{equation} \label{case1:sr2_kdip}
	k_{\rm dip}	\simeq {\sqrt{5} \over 2} {(-\tau_{2})^{3/2} \over (-\tau_{1})^{5/2}} ~,
\end{equation}
or equivalently,
\begin{equation} \label{case1:sr2_kdip_2}
	(- k_{\rm dip} \tau_{1})^{2} \simeq {5 \over 4} e^{-3 N_{\rm USR}} ~.
\end{equation}
where $N_{\rm USR} \simeq \ln(\tau_{1}/\tau_{2})$ is the e-folding number of the intermediate USR phase.
The second derivative at $k_{\rm dip}$ is positive, meaning $k_{\rm dip}$ refers to a minimum, i.e., the dip position of the final power spectrum $\mathcal{P}_{\chi}^{\rm SR2}(\tau_{\rm end}, k)$ for fixed $\tau_{1,2}$, as shown in Fig.~\ref{fig:case1k}. 
This dip scale $k_{\rm dip}$ arises from the cancellation between the terms $- {8 \over 5} {(-\tau_{1})^{5} \over (-\tau_{2})^{3}} k^2$ and ${16 \over 25} {(-\tau_{1})^{10} \over (-\tau_{2})^{6}} k^4 $ in Eq.~\eqref{case1:sr2_super_end}. This cancellation, in turn, stems from the interplay between the positive $(-\tau)^{-6}$ term and negative $(-\tau)^{-3}$ term of the USR power spectrum~\eqref{case1:usr_super}.

It is straightforward to verify that the analytical expressions for the dip position given above in Eqs.~\eqref{case1:sr2_kdip} and~\eqref{case1:sr2_kdip_2} are consistent with the existing literature. Our advance is to provide the first simple analytical forms for these quantities, offering a novel analytical toolkit for theoretical work.
From Eq.~\eqref{case1:sr2_kdip} and the final power spectrum peak that we will discuss in Eq.~\eqref{case2:Ppk}, we derive the relationship between the dip position and the peak value as,
\begin{equation} \label{case1:kdip_Ppk}
	- k_{\rm dip} \tau_{1} \propto \l( \mathcal{P}_{\chi,{\rm pk}} \r)^{-1/4} ~,
\end{equation}
which matches with Eq.~(3.9) given in Ref.~\cite{Tasinato:2020vdk}. For an enhancement of the power spectrum peak by approximately $10^{7}$ (c.f., Eq.~\eqref{case2:gamma_pk} with $N_{\rm USR} \simeq 2.36$), Eq.~\eqref{case1:sr2_kdip_2} yields $k/k_{1} \simeq 0.0324$, where $k_{1} \equiv 1/(-\tau_{1})$. This is consistent with the result of $0.0367$ reported in Ref.~\cite{Fujita:2025imc}. Finally, based on Eq.~\eqref{case1:sr2_super_end}, another relationship is obtained,
\begin{equation} \label{case1:relation2}
	k_{\rm dip}^2 \simeq - {k_{1}^2 \over A_2} ~,
\end{equation}
where $A_2 = - {4 \over 5} {(-\tau_{1})^{3} \over (-\tau_{2})^{3}}$ is the coefficient of $k^2$ term in Eq.~\eqref{case1:sr2_super_end}, as defined in Eq.~(5.14) of Ref.~\cite{Briaud:2025hra}. Equation~\eqref{case1:relation2} agrees with Eq.~(5.16) in Ref.~\cite{Briaud:2025hra}.
Our results confirm the claim in Reference~\cite{Briaud:2025hra} that Eq.~\eqref{case1:relation2} provides an existence criterion for $k_{\rm dip}$, with $A_2<0$ corresponding to the inflation velocity not flipping its sign.

The corresponding dip value can be evaluated from Eq.~\eqref{case1:sr2_super_end} as
\begin{equation} \label{case1:sr2_dipP}
	\mathcal{P}_{\chi,{\rm dip}}
	\equiv 
	\mathcal{P}_{\chi}^{\rm SR2}(\tau_{\rm end}, k_{\rm dip})
	\simeq {5\over2} \mathcal{P}_{\chi,{\rm CMB}}~e^{-3 N_{\rm USR}} ~,
\end{equation}
which is finite and consistent with the numerical results from the linear perturbation theory~\cite{Dimopoulos:2017ged,Byrnes:2018txb, Carrilho:2019oqg,Ozsoy:2019lyy,Tasinato:2020vdk,Cole:2022xqc}. This value originates from the sub-leading term $k^2 (-\tau_{1})^2$ in Eq.~\eqref{case1:sr2_super_end}, while the remaining terms cancel exactly.
Combining the dip value~\eqref{case1:sr2_dipP} and the peak value~\eqref{case2:Ppk}, we derive the relation,
\begin{equation}
	{\mathcal{P}_{\chi,{\rm dip}} \over \mathcal{P}_{\chi,{\rm CMB}}} \propto \l( {\mathcal{P}_{\chi,{\rm pk}} \over \mathcal{P}_{\chi,{\rm CMB}}} \r)^{-1/2} ~,
\end{equation}
which is presented in Eq.~(5.20) of Ref.~\cite{Briaud:2025hra}.

What then sets the upper bound on the attainable enhancement? From the expression~\eqref{case1:sr2_super_end}, the enhancement magnitude scales inversely with $-\tau_{2}$ for given $\tau_{1}$ and $k$. That is, a longer USR phase yields stronger amplification. 
Moreover, Eq.~\eqref{case1:sr2_super_end} shows that $\mathcal{P}_{\chi}^{\rm SR2}(\tau_{\rm end}, k)$ is proportional to $k$. However, for fixed $\tau_{1,2}$ in Case 1, $k$ is bounded by $- k \tau_{2} < - k \tau_{1} \ll 1$. 
Hence, the peak position $k_{\rm pk}$ of the final power spectrum $\mathcal{P}_{\chi}^{\rm SR2}(\tau_{\rm end}, k)$ must arise in other Cases of Fig.~\ref{fig:k_regimes}.

\subsection{Case 3}

Now, we consider Case 3 of Fig.~\ref{fig:k_regimes}, such that the $k$ mode exists the horizon during the last SR phase, meaning it remains subhorizon throughout both the first SR phase and the subsequent USR phase. Thus, we have the relationship $- k \tau_{1} > - k \tau_{2} \gg 1$ for Case 3.

For the first SR phase, $d=3/2$, $A = -H \sqrt{\pi}/2$ and $B=0$, and we have $X_{3/2,\pm} = 0$, Eq.~\eqref{eq:sr_sub_AB} therefore gives
\begin{equation} \label{case3:sr1_sub}
	\lim_{-k\tau \gg 1} \mathcal{P}_{\chi}^{\rm SR1}(\tau, k)
	\simeq {1\over\pi^2} k^{2} (-\tau)^{2} X_{\rm non}^{\rm SR1}(k)
	\simeq \mathcal{P}_{\chi,{\rm CMB}} (-k\tau)^2 ~,
\end{equation}
which matches with Eq.~\eqref{eq:sr1_sub}.

In the subsequent USR phase, we derive the following asymptotic expansions according to Eq.~\eqref{eq:A2B2},
\begin{subequations} \label{case3:usr_sub_coeff}
	\begin{align}
		\lim_{-k\tau_{1} \gg 1} \l( |A_{2}|^2 + |B_{2}|^2 \r) &\simeq {H^2 \pi (-\tau_{1})^{6} \over 4} ~,
		\\
		\lim_{-k\tau_{1} \gg 1} \l(  A_{2}^* B_{2} + A_{2} B_{2}^*\r) &\simeq {3 H^2 \pi (-\tau_{1})^{5} \sin(- 2 k \tau_{1}) \over 4 k} ~,
		\\
		\lim_{-k\tau_{1} \gg 1} i \l( A_{2}^* B_{2} - A_{2} B_{2}^* \r) &\simeq {3 H^2 \pi (-\tau_{1})^{5} \cos( 2 k \tau_{1}) \over 4 k} ~.
	\end{align}
\end{subequations}
These imply that the amplitude $X_{-3/2,\pm}$ of oscillatory terms in Eq.~\eqref{eq:sr_sub_AB} is negligible relative to the non-oscillating amplitude $X_{\rm non}(k)$ for $- k \tau_{1} \gg 1$.
By substituting coefficients in Eq.~\eqref{case3:usr_sub_coeff} into the asymptotic expression~\eqref{eq:usr_sub_AB}, we derive
\begin{equation} \label{case3:usr_sub}
	\lim_{-k\tau \gg 1} \mathcal{P}_{\chi}^{\rm USR}(\tau, k)
	\simeq \mathcal{P}_{\chi,{\rm CMB}} (-\tau_{1})^{6} k^2 (-\tau)^{-4} ~,
\end{equation}
which indicates that the USR power spectrum on subhorizon scales exhibits a growing mode $k^2 (-\tau)^{-4}$. At the transition $\tau_{1}$, Eq.~\eqref{case3:usr_sub} gives
\begin{equation}
	\lim_{-k\tau \gg 1} \mathcal{P}_{\chi}^{\rm USR}(\tau_{1}, k)
	\simeq \mathcal{P}_{\chi,{\rm CMB}} (-k\tau_{1})^{2} ~,
\end{equation}
which matches the preceding SR result~\eqref{case3:sr1_sub}, as expected from the junction condition~\eqref{eq:junction_condition}.

\begin{figure*}[h]
	\centering
	\includegraphics[width=0.33\textheight]{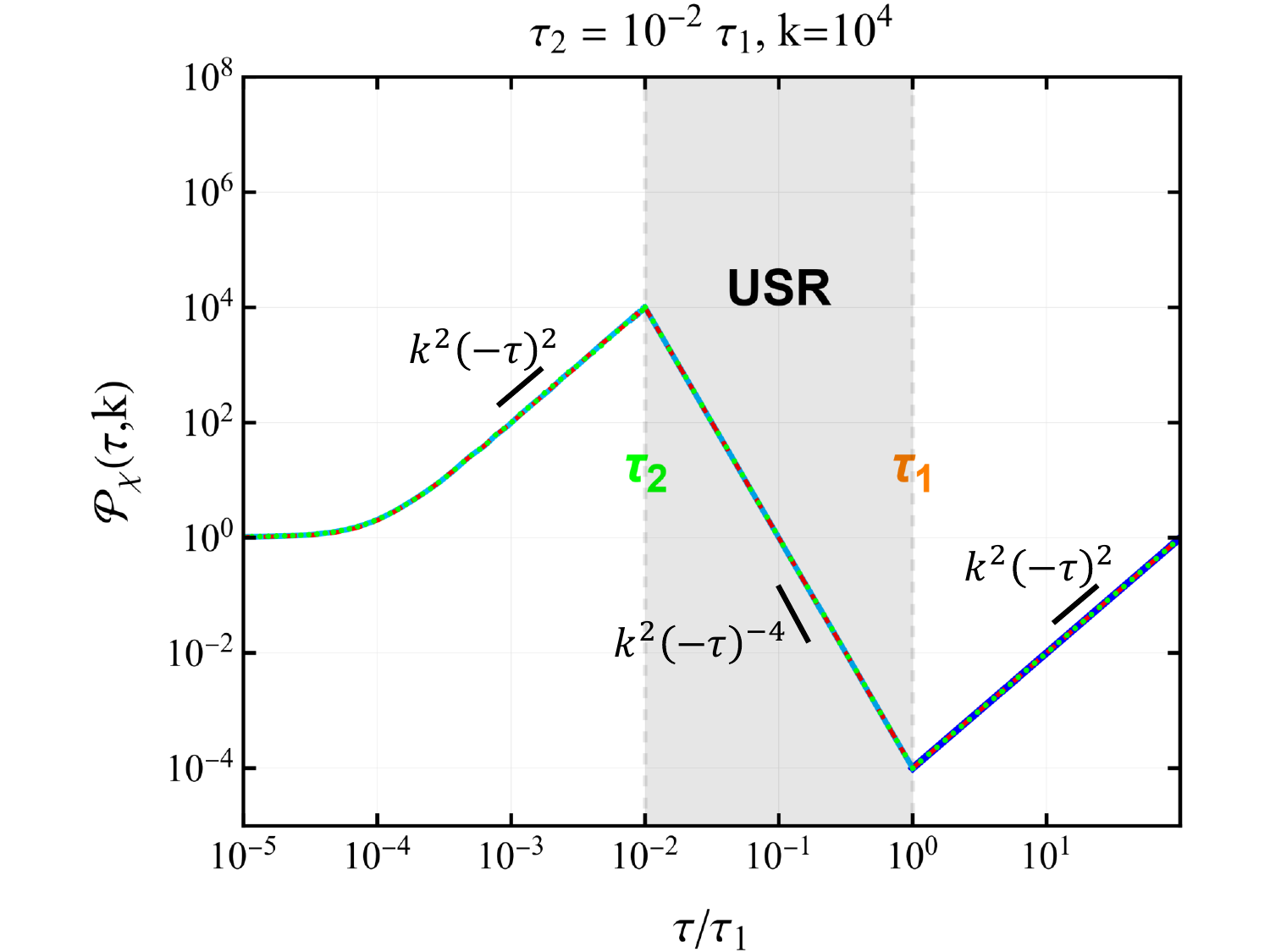}
	\includegraphics[width=0.34\textheight]{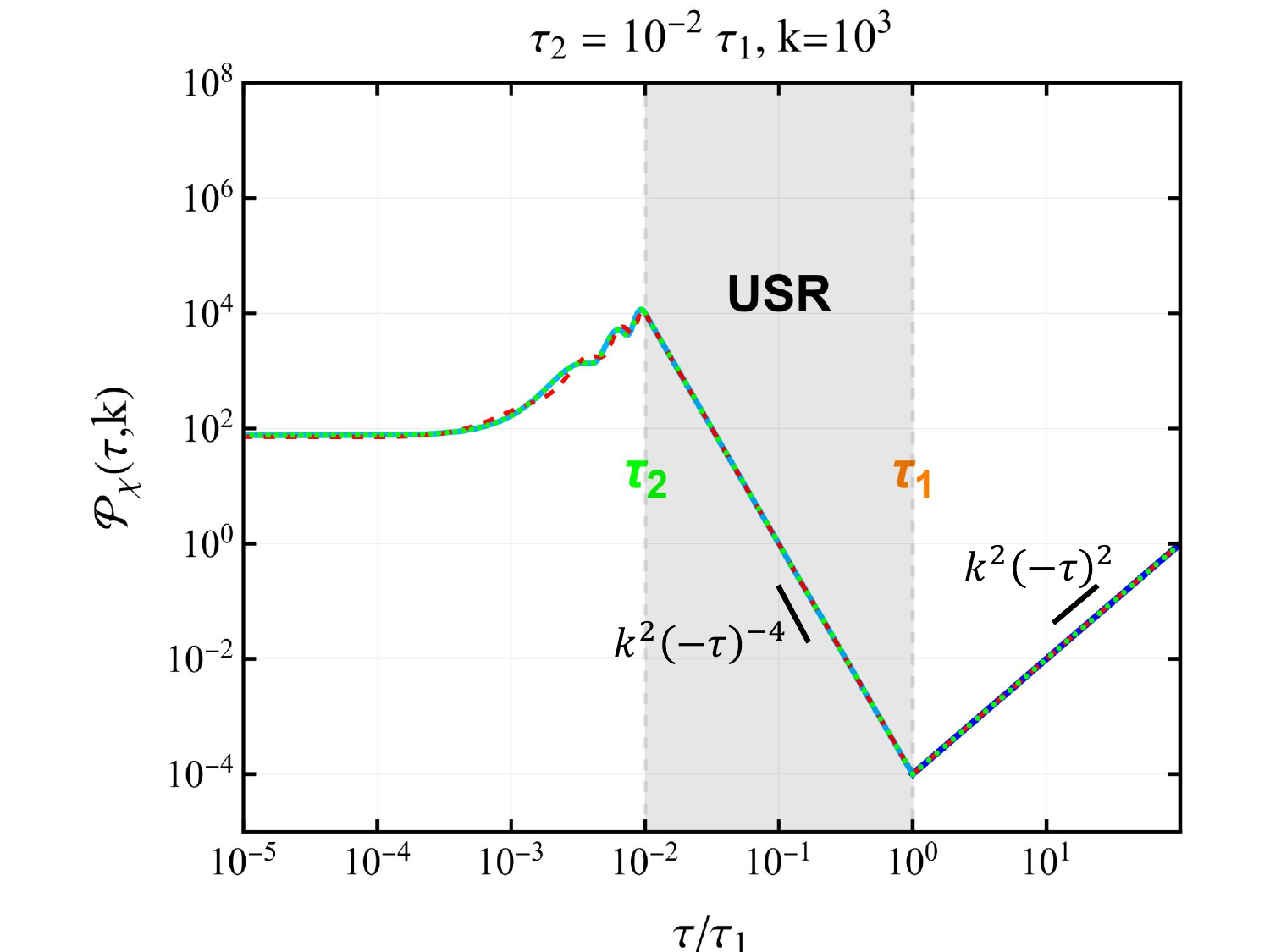}
	\caption{
		The time evolutions of power spectra across SR-USR-SR transitions in Case 3 for $k=10^3$ and $k=10^4$, respectively. 
		The blue solid curves refer to the exact solutions~\eqref{eq:chi_sol_AB} with the coefficients~\eqref{eq:A3B3}, while the red dashed curves denote the asymptotic results given in Eqs.~\eqref{case3:sr1_sub}, \eqref{case3:usr_sub} and \eqref{case3:sr2}. 
		The green dots refers to the numerical results of $\mathcal{P}_{\chi}(\tau, k)$ based on the reconstructed SR-USR-SR inflationary potential~\eqref{eq:final_potential}.
		The parameters are chosen as $\tau_{2}= 10^{-2} \tau_{1} = - 0.01$. 
		All results are normalized by $\mathcal{P}_{\chi}^{\rm USR}(\tau_{1}, k)$ based on Eq.~\eqref{eq:chi_sol_AB}.
	}
	\label{fig:case3}
\end{figure*}

During the final SR phase, the mode exits the horizon at a later time. Similar to Sec.~\ref{subsec:case1_sr2}, treating it as a two-parameter problem, we determine the leading-order terms of the coefficients in Eq.~\eqref{eq:A3B3} as
\begin{subequations} \label{case3:sr2_sub_coeff}
	\begin{align}
		\lim_{- k \tau_{1} > - k \tau_{2} \gg 1} \l( |A_{3}|^2 + |B_{3}|^2 \r) 
		&\simeq {H^2 \pi \over 4} {(-\tau_{1})^{6} \over (-\tau_{2})^{6}} ~,
		\\
		\lim_{- k \tau_{1} > - k \tau_{2} \gg 1} \l(  A_{3}^* B_{3} + A_{3} B_{3}^*\r) 
		&\simeq {3 H^2 \pi \over 4} {(-\tau_{1})^{6} \sin(- 2 k \tau_{2}) \over k (-\tau_{2})^{7} } ~,
		\\
		\lim_{- k \tau_{1} > - k \tau_{2} \gg 1} i \l( A_{3}^* B_{3} - A_{3} B_{3}^* \r) 
		&\simeq {3 H^2 \pi \over 4} { (-\tau_{1})^{6} \cos(- 2 k \tau_{2}) \over k (-\tau_{2})^{7} } ~.
	\end{align}
\end{subequations}
If the $k$ mode remains deep inside the horizon at the second transition $\tau_{2}$, such that $- k \tau_{2} \gg 1$, the non-oscillating amplitude $X_{\rm non}^{\rm SR2}(k)$ also dominates over the oscillating amplitude $X_{3/2,\pm}$(c.f., the left panel of Fig.~\ref{fig:case3}). 
Conversely, if the mode exits the horizon shortly after $\tau_{2}$, $X_{3/2,\pm}$ becomes comparable to $X_{\rm non}^{\rm SR2}(k)$, as illustrated in the right panel of Fig.~\ref{fig:case3}. 
Inserting expressions in Eq.~\eqref{case3:sr2_sub_coeff} into the exact solution~\eqref{eq:sr_sub_AB} gives the leading expression, 
\begin{equation} \label{case3:sr2}
		\begin{aligned}
			\mathcal{P}_{\chi}^{\rm SR2}(\tau, k)
			\simeq& 
			{H^2 \over 4\pi^2} {(-\tau_1)^6 \over (-\tau_2)^6}
			\Bigg\{
			1 - 3 {\sin\l[ 2 k (- \tau_{2} + \tau) \r] \over k (-\tau_2) }
			+ \l[ k^2 - 3 k {\sin\l[ 2 k (\tau_{2} + \tau) \r] \over -\tau_2 } \r] (-\tau)^2
			\Bigg\} ~,
		\end{aligned}
\end{equation}
where we have dropped the $(-\tau)$ term in Eq.~\eqref{eq:sr_sub_AB} as it is subdominant: compared to the $(-\tau)^2$ term on subhorizon scales, and to the constant mode (i.e., the third line in Eq.~\eqref{eq:sr_sub_AB}) on superhorizon scales. 
In addition, the non-oscillatory part $X_{\rm non}$ dominates all other oscillating parts in the constant mode on superhorizon scales. 
The red dashed curves in Fig.~\ref{fig:case3} denote the asymptotic expressions given in Eqs.~\eqref{case3:sr1_sub}, \eqref{case3:usr_sub} and~\eqref{case3:sr2}, while the blue solid curves represent the exact solutions based on Eq.~\eqref{eq:chi_sol_AB} with the coefficients~\eqref{eq:A3B3}.

To verify consistency, we apply Eq.~\eqref{case3:sr2} to derive the initial value at the transition $\tau_{2}$,
\begin{equation}
	\mathcal{P}_{\chi}^{\rm SR2}(\tau_{2}, k) 
	\simeq \mathcal{P}_{\chi,{\rm CMB}} {(-\tau_{1})^{6} \over (-\tau_{2})^{4}} k^{2} ~,
\end{equation}
which matches the results derived from Eq.~\eqref{case3:usr_sub}.
According to Eq.~\eqref{case3:sr2}, we calculate the final value as
\begin{equation} \label{case3:chi_end}
		\mathcal{P}_{\chi}^{\rm SR2}(\tau_{\rm end}, k)
		\simeq 
		\mathcal{P}_{\chi,{\rm CMB}} ~e^{6 N_{\rm USR}} \l[ 1 - 3 {\sin\l[ 2 k (- \tau_{2} + \tau_{\rm end}) \r] \over k (-\tau_2) } \r] ~.
\end{equation}
Note that the second transition leads to the oscillation with the frequency about $- 2 \tau_{2}$ ($\tau_{\rm end}$ approaches 0), which are shown by the red dashed curve in Fig.~\ref{fig:case2k}. 
{\it Hence, two oscillatoin patterns in the final power spectrum reveals the information of the two transition times $\tau_{1}$ and $\tau_{2}$.}
The first crest of the oscillation gives the maximum of the above power spectrum, is about $2.3$ times $\mathcal{P}_{\chi,{\rm CMB}} ~e^{6 N_{\rm USR}}$, which is consistent with Fig.~5 in Ref.~\cite{Pi:2022zxs}. This oscillation arises from the second transition USR-SR, which has also been observed in Refs.~\cite{Byrnes:2018txb,Pi:2022zxs}. In the large-$k$ regime, Eq.~\eqref{case3:chi_end} gives the vanilla value~\cite{Byrnes:2018txb}
\begin{equation} \label{case3:chi_end_large}
		\mathcal{P}_{\chi}^{\rm SR2}(\tau_{\rm end}, k_{\rm large})
		\simeq
		\mathcal{P}_{\chi,{\rm CMB}} ~e^{6 N_{\rm USR}} ~.
\end{equation}

We introduce a enhancement factor to characterise the enhancement with respect to the vanilla superhorizon SR result~\eqref{eq:sr1_super},
\begin{equation} \label{eq:enhancement_factor}
	\gamma \equiv { \mathcal{P}_{\chi}^{\rm SR2}(\tau_{\rm end}, k_{\rm large}) \over \mathcal{P}_{\chi,{\rm CMB}} } ~.
\end{equation}
Equation~\eqref{case3:chi_end} yields
\begin{equation} \label{case3:factor}
	\gamma \simeq e^{6 N_{\rm USR}} ~,
\end{equation}
which is consistent with the conventional results of USR inflation~\cite{Dimopoulos:2017ged,Byrnes:2018txb,Carrilho:2019oqg}.
As expected, the longer USR duration leads to larger enhancement.
The final value~\eqref{case3:chi_end_large} is independent of $k$, this scale-invariant enhancement happens for all modes satisfying $k \gg 1/(-\tau_{2})$. Because those modes are on superhorizon scales during the last phase. This fact is also evident in the numerical results, e.g., Fig.~2 in \cite{Byrnes:2018txb} and Fig.~in Ref.~\cite{Carrilho:2019oqg}.

\subsection{Case 2}

Now, we consider Case 2 of Fig.~\ref{fig:k_regimes}, such that $- k\tau_{2} < 1 < - k\tau_{1}$, meaning the $k$ mode exits the horizon during the intermediate USR phase.
During the first SR phase, we can directly apply the subhorizon asymptotic power spectrum~\eqref{case3:sr1_sub}. 

\begin{figure*}[h]
	\centering
	\includegraphics[width=0.32\textheight]{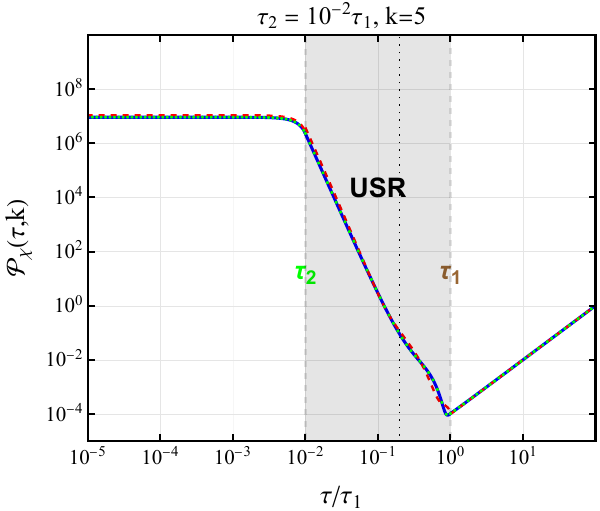}
	\includegraphics[width=0.32\textheight]{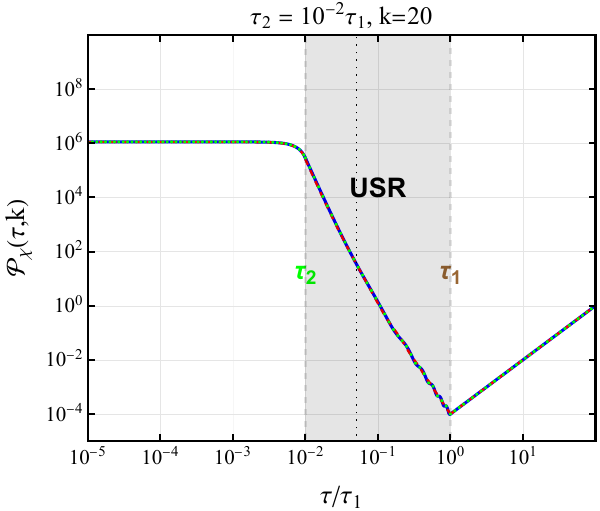}
	\caption{
		The evolutions of power spectra across SR-USR-SR transitions in Case 2 for $k=5$ and $k=20$, respectively. The blue solid curve refers to the exact solution~\eqref{eq:chi_sol_AB} with the coefficients~\eqref{eq:A3B3}, while the red dashed curve denotes the asymptotic expressions given in Eqs.~\eqref{case3:sr1_sub}, \eqref{case2:usr} and \eqref{case2:sr2}. 
		The green dots refers to the numerical results of $\mathcal{P}_{\chi}(\tau, k)$ based on the reconstructed SR-USR-SR inflationary potential~\eqref{eq:final_potential}.
		The parameters are chosen as $\tau_{2}= 10^{-2} \tau_{1} = - 0.01$.
		All results are normalized by $\mathcal{P}_{\chi}^{\rm USR}(\tau_{1}, k)$ based on Eq.~\eqref{eq:chi_sol_AB}.
	}
	\label{fig:case2}
\end{figure*}

The evolution of a mode from subhorizon to superhorizon scales during the USR phase involves complexity, we therefore employ the exact result~\eqref{eq:usr_sub_AB} to preserve accuracy.
For $- k\tau_{1} > 1$, we substitute Eq.~\eqref{case3:usr_sub_coeff} into Eq.~\eqref{eq:usr_sub_AB} and obtain the leading-order asymptotics according to three Rules in Introduction,
\begin{equation} \label{case2:usr}
	\begin{aligned}
		&\mathcal{P}_{\chi}^{\rm USR}(\tau, k)
		\simeq
		\mathcal{P}_{\chi,{\rm CMB}} (-\tau_{1})^{6} 
		\times
		\Bigg[ k^{2} (-\tau)^{-4}
		+ {6\cos[2k(\tau - \tau_1)] \over -\tau_{1}} (-\tau)^{-5}
		+ (-\tau)^{-6}
		\Bigg] ~,
	\end{aligned}
\end{equation}
which possesses three growing modes, scaling as $(-\tau)^{-4}$, $(-\tau)^{-5}$ and $(-\tau)^{-6}$. 
At the transition time $\tau_{1}$, we have $(-\tau)^{-4} >(-\tau)^{-5} > (-\tau)^{-6}$ for $\sqrt{6} < - k\tau_{1} < \tau_{1}/\tau_{2}$, while $(-\tau)^{-5} > (-\tau)^{-4} > (-\tau)^{-6}$ for $1 < - k\tau_{1} < \sqrt{6}$. If the ratio $\tau_{1}/\tau_{2}$ does not exceed $\sqrt{6}$, only the later case applies.
\footnote{However, the asymptotic power spectrum~\eqref{case2:usr} dose not yield a very robust approximation for $1 < - k\tau_{1} < \sqrt{6}$, since $- k\tau_{1}$ not sufficiently large compared to unity. In such a case, retaining more terms in the coefficients of Eq.~\eqref{case3:usr_sub_coeff} becomes necessary and no simple form is attainable. }
Note that the $(-\tau)^{-5}$ term is also oscillating in $\tau$ on subhorizon scales ($- k \tau > 1$), the oscillation frequency is proportional to $k$ as shown in both panels of Fig.~\ref{fig:case2}. In these two panels, $\mathcal{P}_{\chi}^{\rm USR}(\tau, k)$ first grows as $(-\tau)^{-4}$ and oscillates, then switches to a $(-\tau)^{-6}$ scaling near the horizon-crossing $\tau_{*}=-1/k$, marked by the black dotted vertical lines.

For the final SR phase, taking $y < 1 < x$, we obtain the following asymptotic expansions based on the coefficients in Eq.~\eqref{eq:A3B3},
\begin{subequations} \label{case2:sr2_coeff}
		\begin{align}
			\lim_{- k \tau_2 < 1 < - k \tau_1} |A_{3}-B_{3}|^2 
			&\simeq
			\pi H^2 {(-\tau_{1})^{6} \over (-\tau_{2})^{6}} \l[ 1 + 3 {\sin(- 2 k \tau_{1}) \over (-k\tau_{1})} \r] ~,
			\\
			\lim_{- k \tau_2 < 1 < - k \tau_1} i (A_{3}^* B_{3} - A_{3} B_{3}^* )
			&\simeq
			- {3 \pi H^2 \over 2} k^{-3} {(-\tau_{1})^{6} \over (-\tau_{2})^{9} } \l[ 1 + 3 { \sin(- 2 k \tau_{1}) \over (-k\tau_{1}) } \r] ~,
			\\
			\lim_{- k \tau_2 < 1 < - k \tau_1} (A_{3}^* B_{3} + A_{3} B_{3}^*)
			&\simeq
			{9 \pi  H^2 \over 8} k^{-6} {(-\tau_{1})^{6} \over (-\tau_{2})^{12} } \l[ 1 + 3 {\sin(- 2 k \tau_{1}) \over (-k\tau_{1}) } \r] ~.
		\end{align}
\end{subequations}
Note that the term inside all square brackets, $\l[ 1 + 3 {\sin(- 2 k \tau_{1}) \over (-k\tau_{1})} \r]$, is a common factor in each coefficients. 
\footnote{However, this factor can become negative within a narrow range of $1.9 \lesssim - k\tau_{1} \lesssim 2.6$, which is problematic as it in principle cannot serve as the asymptotic form for an non-negative squared norm $|A_{3}-B_{3}|^2$. This is because the asymptotic results~\eqref{case2:sr2_coeff} are not robust for small $x$ and large $y$ within the regime $y < 1 < x$. Despite this, it is found to be harmless for the subsequent discussions, and we are mainly interested in the regimes of $x > 3$.}
Specially, the leading oscillating piece ${ \sin(- 2 k \tau_{1}) \over (-k\tau_{1}) }$ depends only on the variables $- k \tau_{1}$. 
Since the mode is subhorizon at $\tau_{1}$ (i.e., $x = - k\tau_{1} > 1$), we keep the full oscillatory dependence on $x$ in $A_{3}$ and $B_{3}$ (c.f., Eq.~\eqref{eq:A3B3xy}). However, the condition $y < 1$ allows us to apply a Taylor expansion of oscillatory terms in $y$, thereby eliminating all $y$-oscillations in Eq.~\eqref{case2:sr2_coeff}.

Plugging those expressions in Eq.~\eqref{case2:sr2_coeff} into the superhorizon asymptotic expression~\eqref{eq:sr_super_AB}, we yield
\begin{equation} \label{case2:sr2}
	\begin{aligned}
		\lim_{- k \tau_2 < 1 < - k \tau_1} \mathcal{P}_{\chi}^{\rm SR2}(\tau, k)
		\simeq
		\mathcal{P}_{\chi,{\rm CMB}} ~e^{6 N_{\rm USR}}
		\times \l[ 1 + 3 { \sin(- 2 k \tau_{1}) \over (-k\tau_{1}) } \r]
		\Bigg[ 4 - 4 {(-\tau)^3 \over (-\tau_{2})^3}
		+ {(-\tau)^6 \over (-\tau_{2})^6} \Bigg] ~.
	\end{aligned}
\end{equation}
At the transition $\tau_{2}$, Eq.~\eqref{case2:sr2} gives
\begin{equation}
	\lim_{-k\tau \ll 1} \mathcal{P}_{\chi}^{\rm SR2}(\tau_{2}, k)
	\simeq
	\mathcal{P}_{\chi,{\rm CMB}} ~e^{6 N_{\rm USR}} ~,
\end{equation}
which is consistent with the result derived from Eq.~\eqref{case2:usr} evaluated at $\tau_{2}$. We note that qualitatively similar time evolutions are also presented in Ref.~\cite{Ballesteros:2020qam} (cf. the right panel of Fig.~7 therein), which align with the behavior shown in our Figs.~\ref{fig:case1}, \ref{fig:case3}, and \ref{fig:case2}.

\begin{figure*}[h]
	\centering
	\includegraphics[width=0.34\textheight]{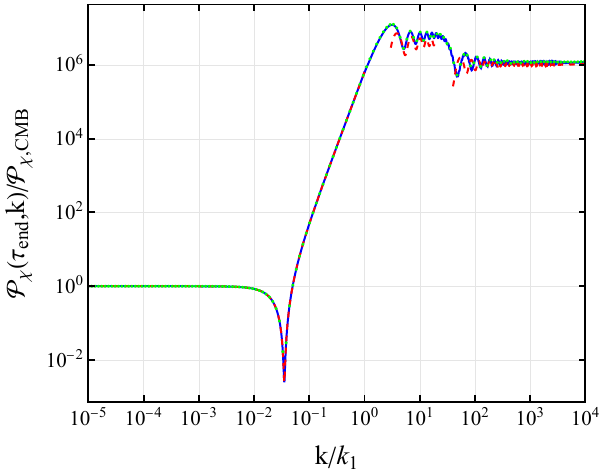}
	\includegraphics[width=0.35\textheight]{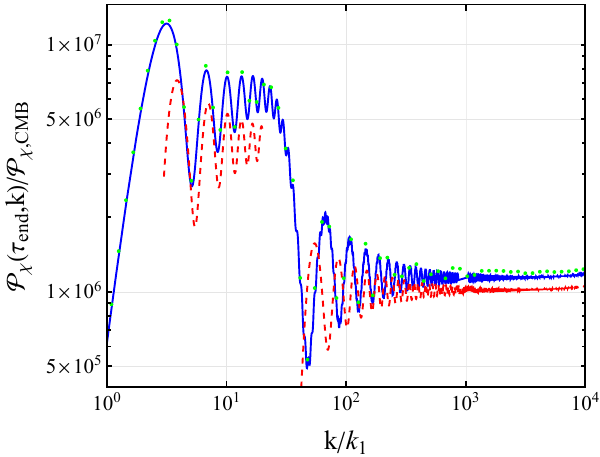}
	\caption{
		The analytical (blue curve), asymptotic (red dashed), and numerical (green dots) results for the final power spectrum $\mathcal{P}_{\chi}(\tau_{\rm end}, k)$.
		The analytical result is obtained from the Hankel function solution~\eqref{eq:chi_sol_AB} with the coefficients given in Eq.~\eqref{eq:A3B3xy}. The asymptotic results combine the expressions provided in Eqs.~\eqref{case1:sr2_super_end}, \eqref{case3:chi_end}, and \eqref{case2:sr2_end}.
		In both panels, the values of $\mathcal{P}_{\chi}(\tau_{\rm end}, k)$ are normalized by $\mathcal{P}_{\chi,{\rm CMB}}$.
	}
	\label{fig:case2k}
\end{figure*}

According to Eq.~\eqref{case2:sr2}, we estimate its value at the end of inflation as
\begin{equation} \label{case2:sr2_end}
	\mathcal{P}_{\chi}^{\rm SR2}(\tau_{\rm end}, k)
	\simeq
	4 \mathcal{P}_{\chi,{\rm CMB}} ~e^{6 N_{\rm USR}} \l[ 1 + 3 { \sin(- 2 k \tau_{1}) \over (-k\tau_{1}) } \r] ~,
\end{equation}
which is plotted in Fig.~\ref{fig:case2k} as a function of $k/k_{1}$. It exhibits an oscillation with an angular frequency $-2\tau_{1}$, which is also reported in Eqs.~(2.12) and (6.5) of Ref.~\cite{Briaud:2025hra}.

The enhancement factor defined in Eq.~\eqref{eq:enhancement_factor} for Eq.~\eqref{case2:sr2_end} is thus calculated as
\begin{equation} \label{case2:factor}
	\gamma
	\simeq
	4 e^{6 N_{\rm USR}} \l[ 1 + 3 { \sin(- 2 k \tau_{1}) \over (-k\tau_{1}) } \r] ~,
\end{equation}
which exceeds the result for Case 3 given in Eq.~\eqref{case3:factor}. These oscillations gradually decay until some wavenumber smaller than $k_{2} = 1/(-\tau_{2})$, see Fig.~\ref{fig:case2k}. We estimate 
\begin{equation} \label{case2:Pk2}
	\mathcal{P}_{\chi}^{\rm SR2}(\tau_{\rm end}, k \rightarrow k_{2})
	\simeq
	4 \mathcal{P}_{\chi,{\rm CMB}} ~e^{6 N_{\rm USR}} ~,
\end{equation}
which agrees with Eq.~\eqref{case3:chi_end} to within one order of magnitude.

Examining Eq.~\eqref{case2:factor}, the maximum enhancement happens around the lower bound $- k_{\rm pk} \tau_{1} = 1$, which determines the peak position,
\begin{equation} \label{case2:kpeak}
	k_{\rm pk} \simeq 4 k_{1} ~,
\end{equation}
meaning $k_{\rm pk}$ exists the horizon soon after the first transition time $\tau_{1}$, which is consistent with previous literature (e.g., Refs.~\cite{Byrnes:2018txb,Carrilho:2019oqg,Briaud:2025hra}). 
The peak value is correspondingly estimated as
\begin{equation} \label{case2:Ppk}
	\mathcal{P}_{\chi,{\rm pk}}
	\equiv
	\mathcal{P}_{\chi}^{\rm SR2}(\tau_{\rm end}, k_{\rm pk})
	\simeq
	7 \mathcal{P}_{\chi,{\rm CMB}} e^{6 N_{\rm USR}} ~,
\end{equation}
which yields the enhancement factor,
\begin{equation} \label{case2:gamma_pk}
	\gamma_{\rm pk} \simeq 7 e^{6 N_{\rm USR}} ~,
\end{equation}
which is approximately $7$ times larger than the value given by Eq.~\eqref{case3:chi_end}, consistent with the numerical result of about $10$~\cite{Briaud:2025hra,Carrilho:2019oqg}.

\section{Conclusion}

In this work, we revisited the behavior of linear comoving curvature perturbations in an inflationary scenario consisting of successive slow-roll, ultra-slow-roll, and slow-roll phases with instantaneous transitions. By combining the junction condition method with carefully constructed asymptotic expansions of Hankel functions, and guided by three simple but powerful rules for identifying dominant contributions across transitions. we have derived accurate analytical expressions for the time evolution of the mode functions and the resulting power spectrum $\mathcal{P}_{\mathcal{R}}(\tau,k)$.

Our analysis provides, for the first time, systematic asymptotic expressions of the linear power spectrum $\mathcal{P}_{\mathcal{R}}(\tau,k)$, which capture the complete time evolution of the spectrum across all relevant $k$ regimes. 
These asymptotic analytical results successfully account for the key features of the USR spectrum: the non-zero dip structure, the characteristic $k^4$ growth, and the two oscillatory patterns observed in the power spectrum at the end of inflation. 
These analytical expressions enhance the predictive power of the model, provide deeper physical insight into perturbation growth during USR phases, and establish a refined analytical basis for confronting future high-precision cosmological observations.

The methodology developed here, i.e., combining the junction method with careful asymptotic analysis of Hankel function solutions, offers a robust framework for studying non-attractor inflationary dynamics more broadly. 
First, our framework can be naturally extended beyond the instantaneous transition limit to scenarios with multiple or smooth transitions~\cite{Cole:2022xqc, Karam:2022nym, Kristiano:2024vst, Briaud:2025hra, Cruces:2026qvl}. 
Such an extension would quantify how transition steepness modulates the dip depth, spectral amplitude and oscillatory features, and would unify the analytical descriptions of instantaneous and smooth transition scenarios under a single scheme.
More broadly, this framework carries over naturally to the general class of non-attractor inflation models, whose dynamics are universally encoded in a time-dependent effective Hubble friction. The instantaneous SR-USR-SR scenario considered here is the standard case: the effective Hubble friction takes piecewise constant values and flips sign abruptly at each transition boundary. Adapting our asymptotic matching scheme to a general value of $d$ in Eq.~\eqref{eq:chi_eom}, or smoothly evolving effective Hubble friction would enable analytical descriptions of perturbation growth and power spectrum evolution for a much wider range of non-attractor scenarios, far beyond the standard USR paradigm.
Second, it would be valuable to extend this linear-order analysis to higher-order perturbation theory, to compute primordial non-Gaussianities generated at USR transitions. These higher-order predictions would provide additional observational handles to distinguish USR inflation from other non-attractor scenarios. 
Finally, more precise calculations of the primordial black hole mass function and the induced gravitational wave spectrum can be performed using our expressions, yielding theoretical templates for upcoming CMB and GW experiments. We leave these interesting extensions to future work.

\appendix

\section{Consistent Check}
\label{app:check}

Plugging Eq.~\eqref{eq:simple_hankel_sr} into Eq.~\eqref{eq:chi_sol_AB}, we derive the power spectrum in the SR phase as
\begin{equation} \label{app1:sr}
	\begin{aligned}
		\mathcal{P}_{\chi}^{\rm SR}(\tau, k)
		&= 
		{|A|^2 + |B|^2\over \pi^3} \l[ 1 + k^{2} (-\tau)^2 \r]
		\\&\quad
		+ {A B^*\over \pi^3} e^{- 2 i k \tau} \l[ - 1 + 2 i k (-\tau) + k^{2} (-\tau)^2 \r]
		\\&\quad
		+ {A^* B \over \pi^3} e^{2 i k \tau} \l[ - 1 - 2 i k (-\tau) + k^{2} (-\tau)^2 \r] ~,
	\end{aligned}
\end{equation}
which is an exact expression obtained without resorting to any perturbative expansion.
On superhorizon scales ($- k \tau \ll 1$), the exponentials $e^{- 2 i k \tau}$ and $e^{2 i k \tau}$ are both dominated by unity. Consequently, the leading asymptotic behavior is given by the constant mode, $\lim_{- k \tau \ll 1} \mathcal{P}_{\chi}^{\rm SR}(\tau, k) 
\simeq |A - B|^2 / \pi^3$. 
Keep terms up to order $(-\tau)^6$, we derive
\begin{equation}
	\begin{aligned}
		\lim_{-k\tau \ll 1} \mathcal{P}_{\chi}^{\rm SR}(\tau, k)
		&\simeq
		{|A-B|^2 \over \pi^3}
		+ { 2i (A^* B - A B^*) \over 3 \pi^3} k^3 (-\tau)^3
		+ { 2 (A^* B + A B^*) \over 9 \pi^3} k^6 (-\tau)^6 ~,
	\end{aligned}
\end{equation}
which coincides with Eq.~\eqref{eq:sr_super_AB}.

For the intermediate USR phase, plugging Eq.~\eqref{eq:simple_hankel_usr} into Eq.~\eqref{eq:chi_sol_AB} gives
\begin{equation}
	\begin{aligned} \label{app2:usr}
		\mathcal{P}_{\chi}^{\rm USR}(\tau, k)
		&=
		{|A|^2 + |B|^2\over \pi^3} \l[ (-\tau)^{-6} + k^{2} (-\tau)^{-4} \r]
		\\&\quad
		+ {A B^* \over \pi^3} e^{-2 i k \tau} \l[ (-\tau)^{-6} - 2 i k (-\tau)^{-5} - k^{2} (-\tau)^{-4} \r]
		\\&\quad
		+ {A^* B \over \pi^3}  e^{2 i k \tau} \l[ (-\tau)^{-6} + 2 i k (-\tau)^{-5} - k^{2} (-\tau)^{-4} \r] ~,
	\end{aligned}
\end{equation}
which is an exact expression obtained without resorting to any perturbative expansion.
On superhorizon scales ($- k \tau \ll 1$), in order to retain all the growing modes, it is necessary to keep the higher-order terms in the expansions of $e^{- 2 i k \tau}$ and $e^{2 i k \tau}$.
Thus, taking the leading terms in Eq.~\eqref{app2:usr} gives the approximate expression
\begin{equation}
	\begin{aligned}
		\lim_{- k \tau \ll 1} \mathcal{P}_{\chi}^{\rm USR}(\tau, k)
		&\simeq
		- { 2(A^* B + A B^*) \over 9 \pi^3} k^6
		+ { 2 (A^* B + A B^*) \over 45 \pi^3} k^8 (-\tau)^2
		\\&\quad
		+ {2i \l(A B^* - A^* B \r) \over 3 \pi^3} k^{3} (-\tau)^{-3}
		+ {|A+B|^2 \over \pi^3} (-\tau)^{-6} ~.
	\end{aligned}
\end{equation}
Comparison with Eq.~\eqref{eq:usr_super_AB}, all growing modes and the constant mode agree as expected, but the decaying $(-\tau)^2$ mode differs because higher-order terms in Hankel functions~\eqref{eq:hankel_expansion} were neglected.

\section{Reconstruction of the SR-USR-SR Inflationary Potential and Numerical Calculations}
\label{app:potential}

In this appendix, we apply the Hamilton-Jacobi formalism to reconstruct the SR-USR-SR inflationary potential~\cite{Motohashi:2014ppa, Motohashi:2019rhu, Wang:2024xdl}, and derive the numerical results of the linear power spectra based on this potential.
First, we reconstruct a potential for a general constant-roll phase, and then apply it into the SR and USR phases. Last, we derive the whole inflationary potential for the SR-USR-SR phases with sharp transitions.

\subsection{Constant-Roll phase}

The Hamilton-Jacobi formalism gives the evolution of inflationary background,
\begin{subequations}
	\begin{align} \label{eq:HJ1}
		{\ddd H(\phi) \over \ddd \phi} &=  - {\dot{\phi} \over 2 M_{\rm Pl}^2} ~,
		\\ \label{eq:HJ2}
		V(\phi) &= 3 M_{\rm Pl}^2 H(\phi)^2 - 2 M_{\rm Pl}^4 \l( {\ddd H(\phi) \over \ddd \phi} \r)^2 ~.
	\end{align}
\end{subequations}
These two equations are derived from the Friedmann equations $H^2 = {1 \over 3 M_{\rm Pl}^2} \l[ \frac12 \dot{\phi}^2 + V(\phi) \r]$ and $\dot{H} = - {1 \over 2 M_p^2} \dot{\phi}^2$ under the assumption $\dot{\phi} \neq 0$.
A shift symmetry $H \rightarrow H+ H_{0}$ is permitted by Eq.~\eqref{eq:HJ1}, and this symmetry leaves $\dot{\phi}$ unchanged. Through Eq.~\eqref{eq:HJ2}, the same shift induces a change in the potential, $V \rightarrow V + 3 M_{\rm Pl}^2 ( H_{0}^2 + 2 H H_{0} )$, reflecting the freedom to choose the zero of potential arbitrarily.

To reconstruct the inflationary potential, we have to solve Eq.~\eqref{eq:HJ1} first. 
Using the slow-roll parameter $\epsilon \equiv - \dot{H}/H^2 = \dot{\phi}^2/( 2M_{\rm Pl}^2 H^2)$ and the convention $\dot{\phi} < 0$, we obtain $\dot{\phi} = - \sqrt{2 \epsilon} M_{\rm Pl} H$.
Then, Eq.~\eqref{eq:HJ1} can be rewritten as
\begin{equation} \label{eq:hubb_1st}
	{\ddd H(\phi) \over \ddd \phi} - {\sqrt{\epsilon(\phi)/2} \over M_{\rm Pl}} H(\phi) = 0 ~.
\end{equation}
Although this equation does not manifestly exhibit the shift symmetry $H \rightarrow H+ H_{0}$, the symmetry is respected once the slow-roll parameter is transformed as $\epsilon \rightarrow - \dot{H}/(H + H_{0})^2$ under the shift symmetry. 
It is convenient to use the relation $\eta = { \dot{\epsilon} \over H \epsilon }
= 2 {\ddot{\phi} \over H \dot{\phi}} + 2 \epsilon$.

For a constant-roll phase, defined by the constant second slow-roll parameter $\eta = \dot{\epsilon}/( H \epsilon)$, we can derive the equation for $\epsilon(\phi)$ from this definition,
\begin{equation} \label{eq:epsilon}
	{\ddd \epsilon(\phi) \over \ddd \phi} + { \eta \over M_{\rm Pl} } \sqrt{\epsilon(\phi) \over 2} = 0 ~,
\end{equation}
which is solved as
\begin{equation} \label{eq:epsilon_sol}
	\sqrt{\epsilon(\phi) \over 2} = \sqrt{\epsilon_{s} \over 2} - {\eta \over 4} { \phi - \phi_{s} \over M_{\rm Pl}} ~,
\end{equation}
where $\epsilon_{s} \equiv \epsilon(\phi_{s})$ is the slow-roll parameter at some reference field value $\phi_{s}$. Observe that this solution is consistent with the convention $\dot{\phi} < 0$. 
Importantly, the positivity of $\epsilon$ implies that the above solution is only valid for some range of $\phi$, 
\begin{equation}
	{ \phi - \phi_{s} \over M_{\rm Pl}} > {4 \over \eta} \sqrt{\epsilon_{s} \over 2} \quad \text{for} \quad \eta<0 ~.
\end{equation}
However, for the slow-roll phase (i.e., $\eta = 0$) and the positive-$\eta$ phase, it is straightforward to check that no constraint on the field range. Hence, the field range for a constant-roll phase with $\eta < 0$ is given by
\begin{equation} \label{eq:field_range}
	{ \phi_{s} \over M_{\rm Pl} } + {4 \over \eta} \sqrt{\epsilon_{s} \over 2} 
	< { \phi \over M_{\rm Pl} } < { \phi_{s} \over M_{\rm Pl} } ~.
\end{equation}

With the solution~\eqref{eq:epsilon_sol}, Eq.~\eqref{eq:hubb_1st} becomes
\begin{equation} \label{eq:master}
	{\ddd H(\phi) \over \ddd \phi} - {1 \over M_{\rm Pl}} \l( \sqrt{\epsilon_{s} \over 2} - {\eta \over 4} { \phi - \phi_{s} \over M_{\rm Pl}} \r) H(\phi) = 0 ~,
\end{equation}
which is the master equation for solving the Hubble parameter. Its solution is given as
\begin{equation} \label{eq:Hcr_general}
	H(\phi) = h_1 \exp\l( \sqrt{\epsilon_{s} \over 2} {\phi \over M_{\rm Pl}} - {\eta \over 8} {\phi^2 \over M_{\rm Pl}^2} + {\eta \over 4} {\phi \phi_{s} \over M_{\rm Pl}^2} \r) ~.
\end{equation}
For the slow-roll phase, i.e., $\eta=0$, the SR solution is given by
\begin{equation} \label{eq:Hsr_general}
	H_{\rm SR}(\phi) = h_1 \exp\l( \sqrt{\epsilon_{s} \over 2} {\phi \over M_{\rm Pl}} \r) ~.
\end{equation}
Setting the de Sitter approximation $\epsilon_{s} = 0$, the above solution leads to a constant potential, which is insufficient for our subsequent analysis. Therefore, we apply the quasi de Sitter approximation to the SR phase, i.e., $\epsilon$ is a constant.

The unknown coefficient $h_1$ can be determined by some condition. For instance, we fix $H_{s} \equiv H(\phi_{s})$, and obtain
\begin{equation} \label{eq:Hcr_sol_1st}
	H_{\rm CR}(\phi) = H_{s} \exp\l[ \sqrt{\epsilon_{s} \over 2} { \phi - \phi_{s} \over M_{\rm Pl}} - {\eta \over 8} \l( {\phi - \phi_{s} \over M_{\rm Pl}} \r)^2 \r] ~,
\end{equation}
and
\begin{equation} \label{eq:Hsr_sol_1st}
	H_{\rm SR}(\phi) = H_{s} \exp\l( \sqrt{\epsilon_{s} \over 2} { \phi - \phi_{s} \over M_{\rm Pl}} \r) ~.
\end{equation}
The corresponding potentials can be reconstructed from the Hamilton-Jacobi formalism~\eqref{eq:HJ2}, we obtain
\begin{equation}
	\begin{aligned}
		V_{\rm CR}(\phi) &= V_{s} \exp\l[ \sqrt{2\epsilon_{s}} { \phi - \phi_{s} \over M_{\rm Pl}} - {\eta \over 4} \l( {\phi - \phi_{s} \over M_{\rm Pl}} \r)^2 \r]
		\\&\quad
		\times \l[ 1 + {\eta \over 3 - \epsilon_{s} } \sqrt{\epsilon_{s} \over 2} { \phi - \phi_{s} \over M_{\rm Pl}} - {\eta^2 \over 8(3 - \epsilon_{s})} \l( {\phi - \phi_{s} \over M_{\rm Pl}} \r)^2 \r] ~,
	\end{aligned}
\end{equation}
and
\begin{equation} \label{eq:Vsr}
	V_{\rm SR}(\phi) = V_{s} \exp\l( \sqrt{2\epsilon_{s}} { \phi - \phi_{s} \over M_{\rm Pl}} \r) ~,
\end{equation}
where $V_{s} \equiv V_{\rm SR}(\phi_{s}) = H_{s}^2 M_{\rm Pl}^2 (3 - \epsilon_{s})$. Setting $\eta = -6$, we derive
\begin{equation}
	\begin{aligned}
		V_{\rm USR}(\phi) &= V_{s} \exp\l[ \sqrt{2\epsilon_{s}} { \phi - \phi_{s} \over M_{\rm Pl}} + {3 \over 2} \l( {\phi - \phi_{s} \over M_{\rm Pl}} \r)^2 \r]
		\\&\quad
		\times \l[ 1 - {6 \over 3 - \epsilon_{s} } \sqrt{\epsilon_{s} \over 2} { \phi - \phi_{s} \over M_{\rm Pl}} - {9 \over 2(3 - \epsilon_{s})} \l( {\phi - \phi_{s} \over M_{\rm Pl}} \r)^2 \r] ~,
	\end{aligned}
\end{equation}

\subsection{SR-CR junction}

For the first transition, SR $\rightarrow$ CR, we fix the transition at $\phi_{s}$ and the junction condition, $H_{\rm SR}(\phi_{s}) = H_{\rm CR}(\phi_{s})$. 
Note that, unlike some previous studies (e.g., Refs.~\cite{Byrnes:2018txb, Escriva:2025ftp}), which set the potential during the USR phase to a costant value, the potential for USR phase is not exactly flat, but exhibits a slight tilt, which deviates from the conventional understanding of the USR. This behavior can be explained using the Klein-Gordon equation for the inflaton, which takes the form,
\begin{equation}
	{\ddd V(\phi) \over \ddd\phi} = - \l({1\over2} \eta - \epsilon(\phi) + 3\r) H \dot{\phi} ~.
\end{equation}
As a result, $\ddd V(\phi)/ \ddd\phi > 0$ for the SR phase (where $\epsilon > 0$ and $\eta = 0$), while $\ddd V(\phi) / \ddd\phi < 0$ (where $\epsilon > 0$ and $\eta = -6$) for the USR phase.

Moreover, the potential is only continuous at the transition $\phi_{s}$, but not smooth. This flaw maybe solved using the shift symmetry of the Hubble parameter. In such, it is only necessary to require the the first derivative of $H$ is continuous at the transition. According to Eqs.~\eqref{eq:Hcr_general} and \eqref{eq:Hsr_sol_1st}, we consider the solutions,
\begin{subequations}
	\begin{align}
		H_{\rm SR}(\phi) &= H_{s} \exp\l( \sqrt{\epsilon_{s} \over 2} { \phi - \phi_{s} \over M_{\rm Pl}} \r) ~,
		\\
		H_{\rm CR}(\phi) &= h_1 \exp\l( \sqrt{\epsilon_{s} \over 2} {\phi \over M_{\rm Pl}} - {\eta \over 8} {\phi^2 \over M_{\rm Pl}^2} + {\eta \over 4} {\phi \phi_{s} \over M_{\rm Pl}^2} \r) + H_0 ~,
	\end{align}
\end{subequations}
where we have introduced a constant shift $H_{0}$ to $H_{\rm CR}$. Then, imposing the junction condition $H_{\rm SR}'(\phi_{s}) = H_{\rm CR}'(\phi_{s})$, we solve
\begin{equation}
	H_{\rm CR}(\phi) = H_0 + H_{s} \exp\l[ \sqrt{\epsilon_{s} \over 2} { \phi - \phi_{s} \over M_{\rm Pl}} - {\eta \over 8} \l( {\phi - \phi_{s} \over M_{\rm Pl}} \r)^2 \r] ~.
\end{equation}
Next, we impose the junction condition $H_{\rm SR}(\phi_{s}) = H_{\rm CR}(\phi_{s})$ to solve $H_{0}$, and it shows that
\begin{equation}
	H_{0} = 0 ~.
\end{equation}
This means that the preceding solutions~\eqref{eq:Hcr_sol_1st} and \eqref{eq:Hsr_sol_1st} are already smooth at the transition. 
This behavior stems from the dynamical equation~\eqref{eq:hubb_1st}: the first derivative $H'$ is determined by $\epsilon$ and $H$---both of which are continuous at the transition. $H'$ itself inherits this continuity, ensuring smoothness of the solutions.
However, to ensure the potential is also smooth at the transition, the second derivative of Hubble parameter is must be smooth at the transition, which is not the case when $\eta$ is discontinuous.

Before delving into the smooth junction of $\eta$, we would like to make some comments on the popular method for reconstructing inflationary potential using the second-order equation for the Hubble parameter~\cite{Motohashi:2014ppa, Motohashi:2019rhu}, rather than directly using the first-order equation~\eqref{eq:master}.
Taking the time derivative of the both sides of Eq.~\eqref{eq:HJ1}, we obtain a closed equation for the Hubble parameter as
\begin{equation} \label{eq:hubb_2nd}
	{\ddd^2 H(\phi) \over \ddd \phi^2} + {\beta \over 2 M_{\rm Pl}^2} H(\phi) = 0 ~,
\end{equation}
where $\beta \equiv {\ddot{\phi} \over H \dot{\phi}}$ is a dimensionless constant~\cite{Motohashi:2019rhu}, and it is related to $\epsilon$ and $\eta$ as
\begin{equation} \label{eq:beta_relation}
	\beta(\phi) = {1\over2} \eta - \epsilon(\phi) ~.
\end{equation}
Combining Eqs.~\eqref{eq:epsilon_sol} and \eqref{eq:beta_relation}, Eq.~\eqref{eq:hubb_2nd} has the following general solution,
\begin{equation} \label{eq:hubble_sol_2nd}
	\begin{aligned}
		H(\phi) &= h_1 \exp\l( \sqrt{\epsilon_{s} \over 2} {\phi \over M_{\rm Pl}} - {\eta \over 8} {\phi^2 \over M_{\rm Pl}^2} + {\eta \over 4} {\phi \phi_{s} \over M_{\rm Pl}^2} \r) 
		+ h_2 {\sqrt{\pi} M_{\rm Pl} \over \sqrt{\eta}} \exp\Big[
		- {2\epsilon_{s} \over \eta}
		- \sqrt{2 \epsilon_{s}} { \phi_{s} \over M_{\rm Pl} }
		\\&\quad - {\eta\over4} {\phi_{s}^{2} \over M_{\rm Pl}^{2}}
		+ \sqrt{ {\epsilon_{s} \over 2} } {\phi_{s} \over M_{\rm Pl}}
		- {\eta\over8} {\phi^{2} \over M_{\rm Pl}^{2}}
		+ {\eta\over4} {\phi \phi_{s} \over M_{\rm Pl}^{2}} \Big]
		\times {\rm Erfi}\l[ \sqrt{2 \epsilon_{s} \over \eta} + {\sqrt{\eta}\over2} {\phi_{s} - \phi \over M_{\rm Pl}} \r] ~,
	\end{aligned}
\end{equation}
for $\eta \neq 0$, with the constants $h_{1,2}$ to be determined by two conditions, both $H(\phi_{s})$ and $H'(\phi_{s})$. Here, ${\rm Erfi}(x) \equiv 2/\sqrt{\pi} \int_{0}^{x} e^{t^2} \ddd t$ is the imaginary error function.

For the slow-roll inflation ($\eta = 0$), Eq.~\eqref{eq:hubb_2nd} leads to
\begin{equation}
	H_{\rm SR}(\phi) = h_1 \exp\l( \sqrt{\epsilon_{s} \over 2} {\phi \over M_{\rm Pl}} \r) + h_2 \exp\l( -\sqrt{\epsilon_{s} \over 2} {\phi \over M_{\rm Pl}} \r)  ~.
\end{equation}
Imposing the conditions,
\begin{subequations} \label{eq:2nd_junction}
	\begin{align}
		H_{\rm SR}(\phi_{s}) &= H_{s} ~, 
		\\ \label{eq:hsr_deriv_junction}
		H_{\rm SR}'(\phi_{s}) &= \sqrt{\epsilon_{s} \over 2} { H_{s} \over M_{\rm Pl} } ~.
	\end{align}
\end{subequations}
and we derive the Hubble parameter for the SR phase,
\begin{equation}
	H_{\rm SR}(\phi) = H_{s} \exp\l( \sqrt{\epsilon_{s} \over 2} { \phi - \phi_{s} \over M_{\rm Pl}} \r) ~,
\end{equation}
which matches the solution given by~\eqref{eq:Hsr_sol_1st} as it should be.
Because the junction condition~\eqref{eq:hsr_deriv_junction} is consistent with the first-order equation~\eqref{eq:master}, the solution of the second-order equation necessarily coincides with that of the first-order equation.

Applying the junction conditions, $H_{\rm SR}(\phi_{s})=H_{\rm CR}(\phi_{s})$ and $H_{\rm SR}'(\phi_{s})=H_{\rm CR}'(\phi_{s})$, to the constant-roll phase, we find that the coefficients in $H_{\rm CR}$ are given as
\begin{equation}
	h_1 = \exp\l( - \sqrt{\epsilon_{s} \over 2} {\phi_{s} \over M_{\rm Pl}} - {\eta \over 8} {\phi_{s}^2 \over M_{\rm Pl}^2} \r) ~,
	\quad
	h_2 = 0 ~,
\end{equation}
and CR solution is the same to the solution~\eqref{eq:Hcr_sol_1st}. Similarly, the potentials exhibit the same agreement.

\subsection{SR-USR-SR reconstruction}

For the final SR phase, we have the following solution according to Eq.~\eqref{eq:Hsr_sol_1st},
\begin{equation} \label{eq:hubb_sr2}
	H_{\rm SR2}(\phi) = h_1 \exp\l[ \l( \sqrt{\epsilon_{s} \over 2} - {\eta \over 4} { \phi_{e} - \phi_{s} \over M_{\rm Pl}} \r) {\phi \over M_{\rm Pl}} \r] ~,
\end{equation}
where we have used Eq.~\eqref{eq:epsilon_sol}. 
Applying the junction conditions,
\begin{equation}
	H_{\rm SR2}(\phi_{e}) = H_{\rm CR}(\phi_{e}) ~,
\end{equation}
we can solve 
\begin{equation}
	H_{\rm SR2}(\phi) = H_{s} \exp\l[ \sqrt{\epsilon_{s} \over 2} { \phi - \phi_{s} \over M_{\rm Pl}} + {\eta \over 8} {(\phi_{e} - \phi_{s}) (\phi_{e} + \phi_{s} - 2\phi) \over M_{\rm Pl}^2} \r] ~.
\end{equation}
For the case $\eta < 0$, the field value at the end of the CR phase is bounded from Eq.~\eqref{eq:field_range},
\begin{equation}
	{ \phi_{s} \over M_{\rm Pl} } + {4 \over \eta} \sqrt{\epsilon_{s} \over 2} 
	< { \phi_{e} \over M_{\rm Pl} } < { \phi_{s} \over M_{\rm Pl} } ~.
\end{equation}
It is straightforward to check the smoothness of the Hubble parameter at $\phi_{e}$, the reason follows the same argument given earlier.
The potential can be reconstructed as
\begin{equation}
	\begin{aligned}
		V_{\rm SR2}(\phi) &= V_{s} \exp\l[\sqrt{2\epsilon_{s}} { \phi - \phi_{s} \over M_{\rm Pl}} + {\eta\over4} {(\phi_{e} - \phi_{s}) (\phi_{e} + \phi_{s} - 2\phi) \over M_{\rm Pl}^2} \r]
		\\&\quad
		\times\l[ 1 + {\eta \over 3 - \epsilon_{s}} \sqrt{\epsilon_{s} \over 2} {\phi_{e} - \phi_{s} \over M_{\rm Pl}} - {\eta^2\over8 (3 - \epsilon_{s})} \l( {\phi_{e} - \phi_{s} \over M_{\rm Pl}} \r)^2 \r] ~.
	\end{aligned}
\end{equation}
It is straightforward to check the continuity, i.e., $V_{\rm SR2}(\phi_e) = V_{\rm CR}(\phi_e)$.

\begin{figure*}[ht]
	\centering
	\includegraphics[width=0.3\textheight]{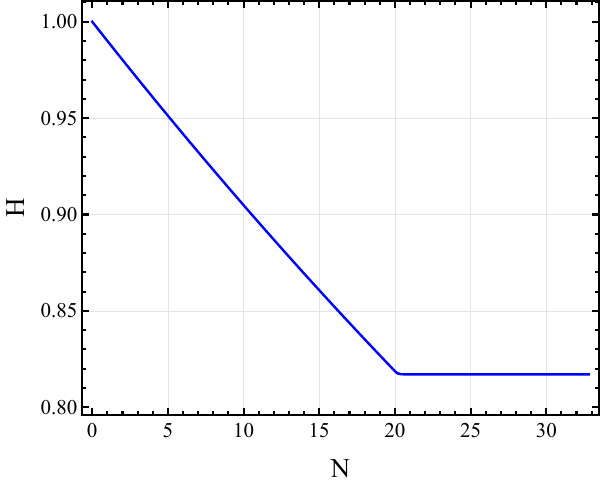}
	\includegraphics[width=0.3\textheight]{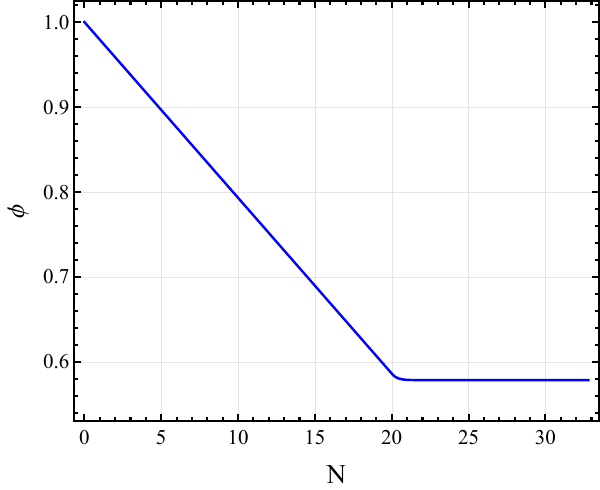}
	\includegraphics[width=0.3\textheight]{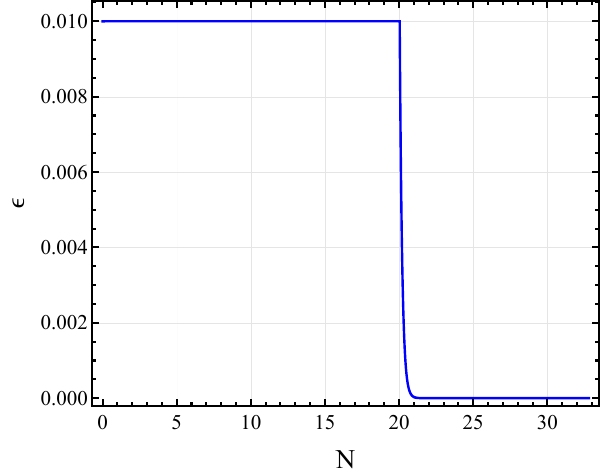}
	\includegraphics[width=0.3\textheight]{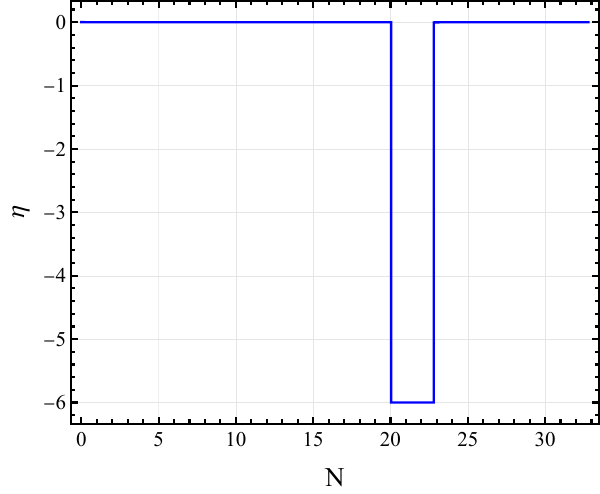}
	\caption{
		The numerical results of (part of) inflationary background evolutions in terms of e-folding number $N$, based on the reconstructed potential~\eqref{eq:final_potential}. The Hubble parameter $H$ and the inflaton field $\phi$ are normalized by values estimated at $N=0$.
	}
	\label{fig:background}
\end{figure*}

With the above preparations, we set $\eta = -6$, and obtain the full expression for the SR-USR-SR inflationary potential,
\begin{equation} \label{eq:final_potential}
	\begin{aligned}
		V_{\rm SR-USR-SR}(\phi) 
		&=
		\l\{
		\begin{matrix}
			&V_{s} \exp\l( \sqrt{2\epsilon_{s}} { \phi - \phi_{s} \over M_{\rm Pl}} \r) ~,  &&{\rm SR1}
			\\
			\\&
			V_{s} \exp\l[ \sqrt{2\epsilon_{s}} { \phi - \phi_{s} \over M_{\rm Pl}} + {3 \over 2} \l( {\phi - \phi_{s} \over M_{\rm Pl}} \r)^2 \r]
			\\&\quad
			\times \l[ 1 - {6 \over 3 - \epsilon_{s} } \sqrt{\epsilon_{s} \over 2} { \phi - \phi_{s} \over M_{\rm Pl}} - {9 \over 2(3 - \epsilon_{s})} \l( {\phi - \phi_{s} \over M_{\rm Pl}} \r)^2 \r] ~,  &&{\rm USR}
			\\
			\\&
			V_{s} \exp\l[\sqrt{2\epsilon_{s}} { \phi - \phi_{s} \over M_{\rm Pl}} + {\eta\over4} {(\phi_{e} - \phi_{s}) (\phi_{e} + \phi_{s} - 2\phi) \over M_{\rm Pl}^2} \r]
			\\&\quad
			\times\l[ 1 + {\eta \over 3 - \epsilon_{s}} \sqrt{\epsilon_{s} \over 2} {\phi_{e} - \phi_{s} \over M_{\rm Pl}} - {\eta^2\over8 (3 - \epsilon_{s})} \l( {\phi_{e} - \phi_{s} \over M_{\rm Pl}} \r)^2 \r] ~,  &&{\rm SR2}
		\end{matrix}
		\r. 
	\end{aligned}
\end{equation}
Based on this piecewise inflationary potential, we numerically solve the master equation~\eqref{eq:eom} to obtain the background evolution shown in Fig.~\ref{fig:background}, as well as the power spectra $\mathcal{P}_{\chi}(\tau, k)$, depicted as green dots in Figs.~\ref{fig:usr_chi_general}–\ref{fig:case1} and \ref{fig:case3}–\ref{fig:case2k}. The numerical strategy for computing $\mathcal{P}_{\chi}(\tau, k)$ is as follows: for each subsequent segment, the integration takes the final values from the preceding segment as its initial conditions, thus ensuring continuity of the dynamics across transition points. This approach has been adopted in previous works, e.g., Refs.~\cite{Wang:2024nmd, Pi:2022zxs, Wang:2024vfv, Wang:2024wxq}.

\section*{ACKNOWLEDGMENT}

This work is supported by National Natural Science Foundation of China (No.12503003, No.1243300), Jiangsu Basic Research Youth Project (No.BK20251001), Postgraduate Research \& Practice Innovation Program of Jiangsu Province (No.KYCX25-4318), National Key R\&D Program of China (No.2021YFC2203100).
	We thank Yudong Luo and Yu-Cheng Qiu for their contributions during the early stages of this work. We thank Vadim Briaud, Alexandros Karam, Niko Koivunen, Eemeli Tomberg, Hardi Veerm\"{a}e for invaluable communications.
	C.C. thanks Diego Cruces and Xiao-Han Ma for their stimulating discussions and valuable feedback on the manuscript.

\bibliographystyle{apsrev4-1}
\bibliography{hubble}
	
\end{document}